\documentclass[trackchanges,twocolumn,twocolappendix]{aastex7}

\usepackage[utf8]{inputenc}
\usepackage[T1]{fontenc}
\usepackage{textcomp}

\begin{document}

\title{ALMA Band 7 Observations of Water Lines in the Protoplanetary Disk of V883 Ori}

\author[orcid=0009-0003-7323-9952,sname='Nakasone']{Hiroto Nakasone}
\affiliation{Department of Astronomy, Graduate School of Science, Kyoto University, Kitashirakawa-Oiwake-cho, Sakyo-ku, Kyoto 606-8502, Japan}
\email[show]{nakasone@kusastro.kyoto-u.ac.jp}

\author[orcid=0000-0003-2493-912X,sname='Notsu']{Shota Notsu}
\affiliation{Department of Earth and Planetary Science, Graduate School of Science, The University of Tokyo, 7-3-1 Hongo, Bunkyo-ku, Tokyo 113-0033, Japan}
\affiliation{Star and Planet Formation Laboratory, Pioneering Research Institute, RIKEN, 2-1 Hirosawa, Wako, Saitama 351-0198, Japan}
\email[]{shota.notsu@eps.s.u-tokyo.ac.jp}

\author[orcid=0000-0001-8002-8473,sname='Yoshida']{Tomohiro C. Yoshida}
\affiliation{National Astronomical Observatory of Japan, 2-21-1 Osawa, Mitaka, Tokyo 181-8588, Japan}
\affiliation{Department of Astronomical Science, The Graduate University for Advanced Studies, SOKENDAI, 2-21-1 Osawa, Mitaka, Tokyo 181-8588, Japan}
\email[]{tomohiroyoshida.astro@gmail.com}  

\author[0000-0002-7058-7682,sname='Nomura']{Hideko Nomura}
\affiliation{National Astronomical Observatory of Japan, 2-21-1 Osawa, Mitaka, Tokyo 181-8588, Japan}
\affiliation{Department of Astronomical Science, The Graduate University for Advanced Studies, SOKENDAI, 2-21-1 Osawa, Mitaka, Tokyo 181-8588, Japan}
\email[]{hideko.nomura@nao.ac.jp}

\author[orcid=0000-0002-6034-2892,sname='Tsukagoshi']{Takashi Tsukagoshi}
\affiliation{Faculty of Engineering, Ashikaga University, Ohmae-cho 268-1, Ashikaga, Tochigi, 326-8558, Japan}
\email[]{takashi.tsukagoshi.astro@gmail.com}

\author[0000-0003-1659-095X]{Tomoya Hirota}
\affiliation{Mizusawa VLBI Observatory, National Astronomical Observatory of Japan, 2-12 Hoshiga-oka, Mizusawa, Oshu-shi, Iwate 023-0861, Japan}
\affiliation{Department of Astronomical Science, The Graduate University for Advanced Studies, SOKENDAI, 2-21-1 Osawa, Mitaka, Tokyo 181-8588, Japan}
\email[]{tomoya.hirota@nao.ac.jp}

\author[0000-0002-6172-9124]{Mitsuhiko Honda}
\affiliation{Faculty of Biosphere-Geosphere Science, Okayama University of Science, 1-1 Ridai-chou, Okayama 700-0005, Japan}
\email[]{hondamt1977@gmail.com}

\author[0000-0002-5082-8880]{Eiji Akiyama}
\affiliation{Division of Fundamental Education and Liberal Arts, Department of Engineering, Niigata Institute of Technology 1719 Fujihashi, Kashiwazaki, Niigata 945-1195, Japan}
\email[]{eakiyama\@niit.ac.jp}

\author[0000-0003-2014-2121]{Alice S. Booth}
\affiliation{Center for Astrophysics, | Harvard \& Smithsonian, 60 Garden St., Cambridge, MA 02138, USA}
\email[]{alice.booth@cfa.harvard.edu}

\author[0000-0003-3119-2087]{Jeong-Eun Lee}
\affiliation{Department of Physics and Astronomy, Seoul National University, 1 Gwanak-ro, Gwanak-gu, Seoul 08826, Republic of Korea}
\affiliation{SNU Astronomy Research Center, Seoul National University, 1 Gwanak-ro, Gwanak-gu, Seoul 08826, Republic of Korea}
\email[]{lee.jeongeun@snu.ac.kr}

\author[0000-0002-0226-9295]{Seokho Lee}
\affiliation{Korea Astronomy and Space Science Institute, 776 Daedeok-daero, Yuseong, Daejeon 34055, Republic of Korea}
\email[]{seokholee@kasi.re.kr}

\begin{abstract}
The FU Orionis star V883 Ori provides a unique opportunity to probe the water snowline in a protoplanetary disk.
During an accretion burst, the enhanced stellar luminosity heats the disk, sublimating ices and bringing volatile species into the gas-phase.
The water snowline, located at $\sim$80 au in the midplane, represents a key boundary for dust growth and volatile delivery to forming planets.
We present Atacama Large Millimeter/submillimeter Array Band 7 observations of V883 Ori that detect two targeted water isotopologue transitions: para-H$_2$$^{18}$O $5_{1,5}$--$4_{2,2}$ at 322 GHz and HDO $3_{3,1}$--$4_{2,2}$ at 335 GHz.
After correcting for Keplerian rotation, we detect HDO and H$_2$$^{18}$O at 23.6$\sigma$ and 9.3$\sigma$, respectively.
Rotational-diagram analysis using a Markov Chain Monte Carlo approach yields $T_\mathrm{rot}=116.89\pm12.81$ K and $N=(4.90\pm1.69)\times10^{15}\,\mathrm{cm}^{-2}$ for H$_2$$^{18}$O, and $T_\mathrm{rot}=87.46\pm4.95$ K and  $N=(4.47\pm0.62)\times10^{15}\,\mathrm{cm}^{-2}$ for HDO.
These results imply water vapor abundances of $N_{\mathrm{H_2O}}/N_{\mathrm{H_2}}\sim3\times10^{-7}$--$5\times10^{-6}$ and an HDO/H$_2$O ratio of $(0.4$--$2.0)\times10^{-3}$ just inside the water snowline, broadly consistent with inheritance from protostellar envelopes.
The HDO line in Band 7 is significantly weaker than predicted from Band 6 extrapolation, showing only $\sim$26\% of the expected strength.
This attenuation can be explained by a more compact, hotter emitting region with an effective radius of $\sim$53 au and/or frequency-dependent dust absorption that enlarges the apparent inner cavity at higher frequency.
Our results highlight both the diagnostic power of water isotopologue lines and the need for higher angular resolution observations to resolve the water snowline and test these scenarios.
\end{abstract}


\section{Introduction}\label{Introduction}
Locating the position of the water snowline, which corresponds to the sublimation front of water molecules \citep[e.g.,][]{1981_hayashi, 1985_hayashi}, is crucial for understanding the evolution of dust grains and the architecture of planetary systems \citep[e.g.,][]{2011_oberg, 2011_oka, 2012_okuzumi, 2019_okuzumi, 2013_ros, 2015_banzatti, 2015_piso, 2016_piso, 2016_cieza, 2017_pinilla, 2017_Schoonenberg, 2021_mori, 2022_notsu, 2023_kondo}.
In addition, the location of the water snowline has implications for the delivery of water to terrestrial planets, offering clues to the origin of Earth's oceans \citep[e.g.,][]{2000_morbidelli, 2012_morbidelli, 2016_morbidelli, 2011_walsh, 2016_ida, 2019_ida, 2016_sato, 2017_raymond, 2021_Lichtenberg, 2023_Banzatti}.

The H$_2$O snowline position depends on the mass accretion rate and luminosity of the central star, and on over the disk lifetime it moves inward as the disk evolves \citep[e.g.,][]{2011_oka, 2015_Harsono, 2021_miley, 2022_Murillo, 2024b_Alarcon}.
Recently, water vapor emission from the inner warm disk and envelopes ($>$ 100 K) of low-mass Class 0--I protostars has been investigated using such as NOEMA (NOrthern Extended Millimeter Array), ALMA (Atacama Large Millimeter/submillimeter Array), $Herschel$ Space Observatory, and James Webb Space Telescope (JWST) \citep[e.g.,][]{2014_persson, 2016_Bjerkeli, 2019_jensen, 2021_jensen, 2020_Harsono, 2021_vanDishoeck, 2025_vanDishoeck, 2023_tobin, 2024_facchini, 2025_leemker}, and HDO emission has also been detected toward the FU Ori object V1057 Cyg \citep{2024_calahan}.
The water vapor abundances and the degree of deuteration of water in the warm region of low-mass Class 0 protostars are also reported, and the HDO/H$_2$O ratios are found to lie roughly between $(6\times10^{-4})$ and $(2\times10^{-3})$ \citep[e.g.,][]{2014_persson, 2019_jensen, 2021_jensen, 2020_Harsono, 2021_Notsu, 2025_vanDishoeck}.

V883 Ori is classified as an FU Orionis (FUor) object located in the Orion A L1641 molecular cloud.
It is currently in a transitional phase between Class I and Class II, characterized by a thin envelope and a massive circumstellar disk \citep{2024_alarcon}.
The source is undergoing a rapid luminosity increase attributed to an accretion burst onto the central protostar.
The mass of the central star has been estimated at approximately 1.3 $\mathrm{M}_\odot$, surrounded by a well-developed Keplerian rotating disk with a mass of at least 0.3 $\mathrm{M}_\odot$ and situated at a distance of 400 pc \citep{2018_kounkel}.
Its bolometric luminosity is about 200 L$_\odot$ \citep{2008_greene, 2016_cieza, 2016_furlan, 2019_lee, 2022_ruiz-rodriguez}.

The elevated luminosity of the central protostar increases the disk temperature, which leads to the thermal desorption of molecules previously frozen onto dust grain surfaces, enabling their detection in the gas phase \citep{2018_vanthoff, 2019_lee, 2023_tobin, 2024_yamato, 2025_leemkera}.
Moreover, since the duration of the accretion outburst ($\sim$100 years) is much shorter than the timescales required for significant chemical processing in the gas-phase \citep[e.g.,][]{2009_nomura}, observations can capture these freshly sublimated molecules from dust grain surfaces before they undergo significant chemical alteration in the gas-phase.
This allows near-direct probing of the disk ice composition.
Based on the intensity break observed in the 0.1 mm continuum emission, \citet{2016_cieza} inferred that the water snowline at the disk midplane lies within approximately 42 au of the central star.
On the disk surface, the water snowline (water snow-surface) may extend out to around 160 au \citep{2018_vanthoff, 2019_lee, 2021_leemker}.
However, more recent ALMA observations of HDO by \citet{2023_tobin} indicate that the water snowline at the midplane is located at a radius of about 80 au.

In addition to water, the detection of complex organic molecules (COMs), whose snowline positions are close to water, represents a key observational challenge in studies of disks.
COMs are of particular interest because they are considered potential precursors of prebiotic species that may have contributed to the origin of life on Earth \citep{2023_ceccarelli}.
Using ALMA Band 7 observations, \citet{2018_vanthoff} reported the first detection of CH$_{3}$OH emission in the V883 Ori disk.
Subsequent studies have identified multiple lines of COMs, including isotopologues, toward this source \citep{2019_lee, 2024_yamato, 2025_jeong, 2025_fadul, 2025_fadula, 2025_Zeng}.
They found that the COMs abundance ratios with respect to methanol are significantly higher than those in the warm protostellar envelopes of IRAS 16293-2422 and similar to the ratios in the solar system comet 67P/Churyumov-Gerasimenko, suggesting the efficient (re)formation of COMs in protoplanetary disks. 
In addition, \citet{2024_yamato} also constrained the $^{12}$C/$^{13}$C and D/H ratios of COMs in protoplanetary disks for the first time (see also \citealt{2025_jeong, 2025_fadul, 2025_Zeng}).
The D/H ratios of methyl formate are slightly lower than the values in IRAS 16293-2422, possibly pointing to the destruction and reformation of COMs in the disk \citep{2024_yamato}. 
As demonstrated by \citet{2023_tobin}, identifying and characterizing water isotopologue emission—particularly determining their intensities and emitting regions—requires careful treatment of potential line blending with nearby contaminated lines, which presents a significant challenge both observationally and in the subsequent analysis.

In this paper, we present new observations of water isotopologue lines toward the V883 Ori disk conducted with ALMA Band 7.
Previous studies have reported detections of H$_2$$^{18}$O at 203 GHz in Band 5 and HDO lines at 225 and 241 GHz in Band 6 \citep{2023_tobin} and a marginal detection of the HDO 335 GHz line in Band 7 was reported by \citet{2019_lee}.
Furthermore, the detection of D$_2$O at 316 GHz in Band 7 has also been reported \citep{2025_leemkera}.
Following up on these works and motivated by the theoretical discussions \citep{2016_notsu, 2017_notsu, 2018_notsu} and previous observational attempts \citep{2019_notsu, 2021_bosman, 2024_facchini}, we report ALMA Band 7 observations aimed at detecting three water transitions—H$_2$$^{16}$O at 321 GHz, H$_2$$^{18}$O at 322 GHz, and HDO at 335 GHz.
Although the H$_2$$^{16}$O line was not detected, we successfully detected the H$_2$$^{18}$O and HDO lines in the V883 Ori disk.
We describe the observational and imaging procedures in Section~\ref{Observations}, present a detailed analysis of the disk-integrated spectra including these detections in Section~\ref{Data Analysis and Result}, and discuss the implications for the physical and chemical structure of the V883 Ori disk as well as the isotopic chemistry of water in protoplanetary disks in Section~\ref{Discussion}.
A summary of our results is provided in Section~\ref{Summary}.

\section{Observations}\label{Observations}
\subsection{ALMA Band 7 Observations}\label{ALMA Band 7 Observations}
V883 Ori was observed in Band 7 with the ALMA program 2021.1.00115.S (Principal Investigator: S.Notsu), targeting the ortho-H$_{2}$$^{16}$O 321.23\,GHz line, the para-H$_{2}$$^{18}$O 322.47\,GHz line, and the HDO 335.40\,GHz line.
The molecular coefficients for the three transitions are reported in Table~\ref{tab:parameters} (see also \citealt{2018_notsu, 2019_notsu}).
We observed the target source on September 7, 2022, using 43 ALMA antennas, in one execution block.
The on-source integration time was 42 minutes, under atmospheric conditions with a precipitable water vapor (PWV) of 0.3 mm.
The baseline lengths ranged from 15.1 m to 783.5 m, yielding an angular resolution of 0\farcs3 and a maximum recoverable scale (MRS) of 4\farcs2.
The MRS is sufficient to recover the V883 Ori disk, whose dust and gas (C$^{18}$O) emission extend to $\sim125$ au ($\sim$0\farcs3) and $\sim320$ au ($\sim$0\farcs8), respectively \citep{2016_cieza}.
For calibration, we used J0423–0120 as the bandpass and amplitude calibrator, and J0607–0834 as the phase calibrator.
Each isotope was observed with an appropriate correlator setup, and the detailed properties of the spectral windows (SPWs) corresponding to each isotope are summarized in Table~\ref{tab:setup}.

\begin{deluxetable}{lcccc}
\tablewidth{0pt}
\tablecaption{Parameters of Our Target Water Lines\label{tab:parameters}}
\tablehead{
\colhead{Isotope} &
\colhead{$J_{K_{a}K_{c}}$} &
\colhead{Frequency} &
\colhead{$A_\mathrm{ul}$} & 
\colhead{$E_\mathrm{u}$}
\\
\colhead{} &
\colhead{} &
\colhead{(GHz)} &
\colhead{(s$^{-1}$)} &
\colhead{(K)}
}
\startdata
\hline
o-H$_2$$^{16}$O &
$10_{29}-9_{36}$ &
321.22568 &
$6.17\times10^{-6}$ &
1862.2
\\
p-H$_2$$^{18}$O &
$5_{15}-4_{22}$  &
322.46517 &
$1.06\times10^{-5}$ &
467.9
\\
HDO &
$3_{31}-4_{22}$  &
335.39550 &
$2.61\times10^{-5}$ &
335.3
\\
\enddata
\end{deluxetable}

\begin{deluxetable}{lcccccc}
\tablewidth{0pt}
\tablecaption{Properties of Image Cubes\label{tab:setup}}
\tablehead{
\colhead{Isotope} &
\colhead{Bandwidth} &
\multicolumn{2}{c}{Channel Width} &
\colhead{Vel. Res.} &
\colhead{Beam Size (P.A.)} &
\colhead{rms}
\\
\colhead{} &
\colhead{(MHz)} &
\colhead{(MHz)} &
\colhead{$\mathrm{(km\,s^{-1})}$} &
\colhead{$\mathrm{(km\,s^{-1})}$} &
\colhead{} &
\colhead{$\mathrm{(mJy\,beam^{-1})}$}
}
\startdata
\hline
o-H$_2$$^{16}$O &
234 &
0.122 &
0.114 &
0.132 &
0\farcs41$\times$0\farcs29 (-75\textdegree) &
5.23
\\
p-H$_2$$^{18}$O &
234 &
0.244 &
0.227 &
0.262 &
0\farcs40$\times$0\farcs31 (-75\textdegree) &
4.25
\\
HDO &
117 &
0.122 &
0.109 &
0.126 &
0\farcs39$\times$0\farcs28 (-75\textdegree)&
4.66
\\
\enddata
\end{deluxetable}

\subsection{Calibration and Imaging}\label{Calibration and Imaging}
Initial calibrations were performed by the ALMA staff using the standard ALMA calibration pipeline version 2022.2.0.64.
Subsequent self-calibration and imaging were performed using the Common Astronomy Software Applications (CASA; \cite{2022_casateam}) version 6.5.3.28.
Before making the line images, we first made the image of the continuum emission using the spw tuned for detecting the continuum emission, in order to perform a hybrid-mapping with self-calibration.
The pipeline-calibrated data were first imaged with the CLEAN algorithm using the CASA task \textit{tclean}.
The robust weighting was used for the imaging with a robustness parameter of 0.5.
We employed the multiscale clean with scales corresponding to 0, 1, 3, 5, and 10 times of the synthesized beam.
After the initial clean image of the continuum emission was created, we performed some rounds of self-calibration and CLEAN imaging (i.e., a hybrid-mapping) for the continuum visibilities using the CASA tasks \texttt{gaincal}, \texttt{applycal}, and \texttt{tclean}.
We conducted 3 rounds of phase-only self-calibration by changing solution intervals from 300 to 6 sec, followed by 1 round of amplitude self-calibration.
The final CLEAN image was thus improved by a factor of 10 with respect to the initial one.
The synthesized beam of the final continuum map was $0\farcs39\times0\farcs30$ with a position angle of -78.3\textdegree, and the rms noise level of the map was 0.28 mJy beam$^{-1}$

By using the calibration table obtained with the self-calibration above, we created the CLEAN images of the target lines.
After applying the calibration table to the visibilities of the target lines with \texttt{applycal}, we performed the continuum subtraction with the task \texttt{uvcontsub} to the calibrated visibilities of each line.
The ranges of the fitspws parameter were set to (0--200, 1600--1919), (0--140, 880--959), and (0--130, 240--330, 610--700) channels for ortho-H$_2$$^{16}$O, para-H$_2$$^{18}$O, and HDO, respectively.
The line images were created using CLEAN from the continuum-subtracted visibilities.
We adopted the briggs weighting with a robust parameter of 0.5.
The multiscale CLEAN was employed with scale parameters corresponding to 0, 1, 3, 5, and 10 times of the synthesized beam of each line map.
The synthesized beamsize and the rms noise level of the CLEAN maps were summarized in Table~\ref{tab:setup}.

\section{Data Analysis and Result}\label{Data Analysis and Result}
\subsection{Spectrum Extraction}\label{Spectrum Extraction}

\begin{figure*}[htb!]
\centering
\includegraphics[width=0.7\textwidth]{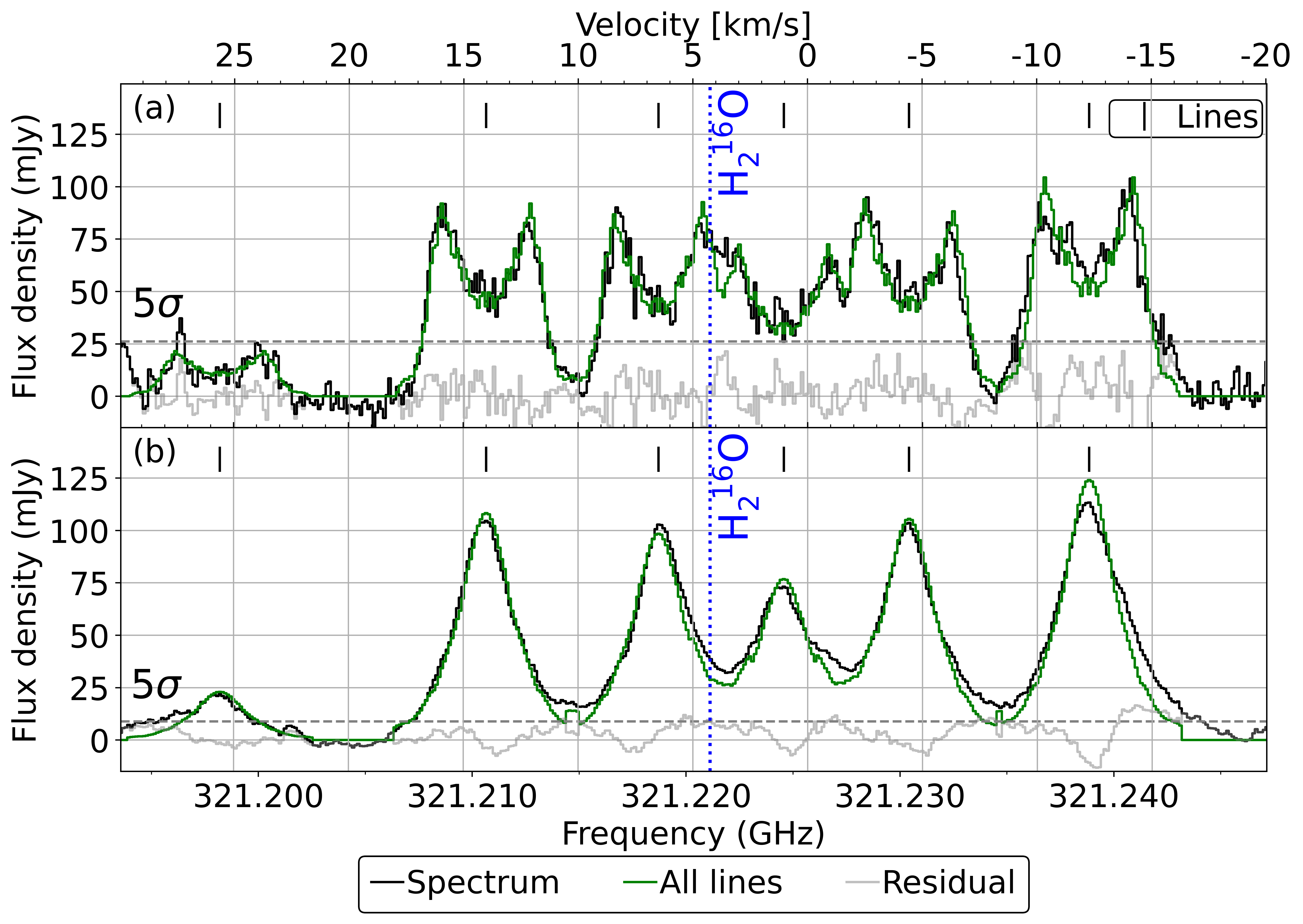}
\caption{\label{fig:spectra_h216o}
(a) Observed spectrum around the H$_{2}$$^{16}$O 321 GHz line (black solid line).
(b) Spectrum after correction for Keplerian rotation (black solid line).
In both (a) and (b), the systemic velocity has not been corrected; the blue dotted lines indicate the source velocity of V883 Ori (4.25 km s$^{-1}$) and mark the expected frequency of the H$_{2}$$^{16}$O transition.
The green solid lines represent the combined fitting models of the nearby contaminated lines.
The gray solid lines show the residuals obtained by subtracting the model spectra from the observed and Keplerian-corrected spectrum.
The vertical black ticks indicate the central frequencies of the nearby contaminated lines (see Line 1-6 in Table~\ref{tab:kep_mask}).
The gray dashed lines indicate the 5$\sigma$ noise level, derived from the standard deviation of spectra taken in an off-source region; (a) from the raw spectrum, and (b) from the Keplerian-rotation-corrected spectrum.}
\end{figure*}

\begin{figure*}[htb!]
\centering
\includegraphics[width=0.7\textwidth]{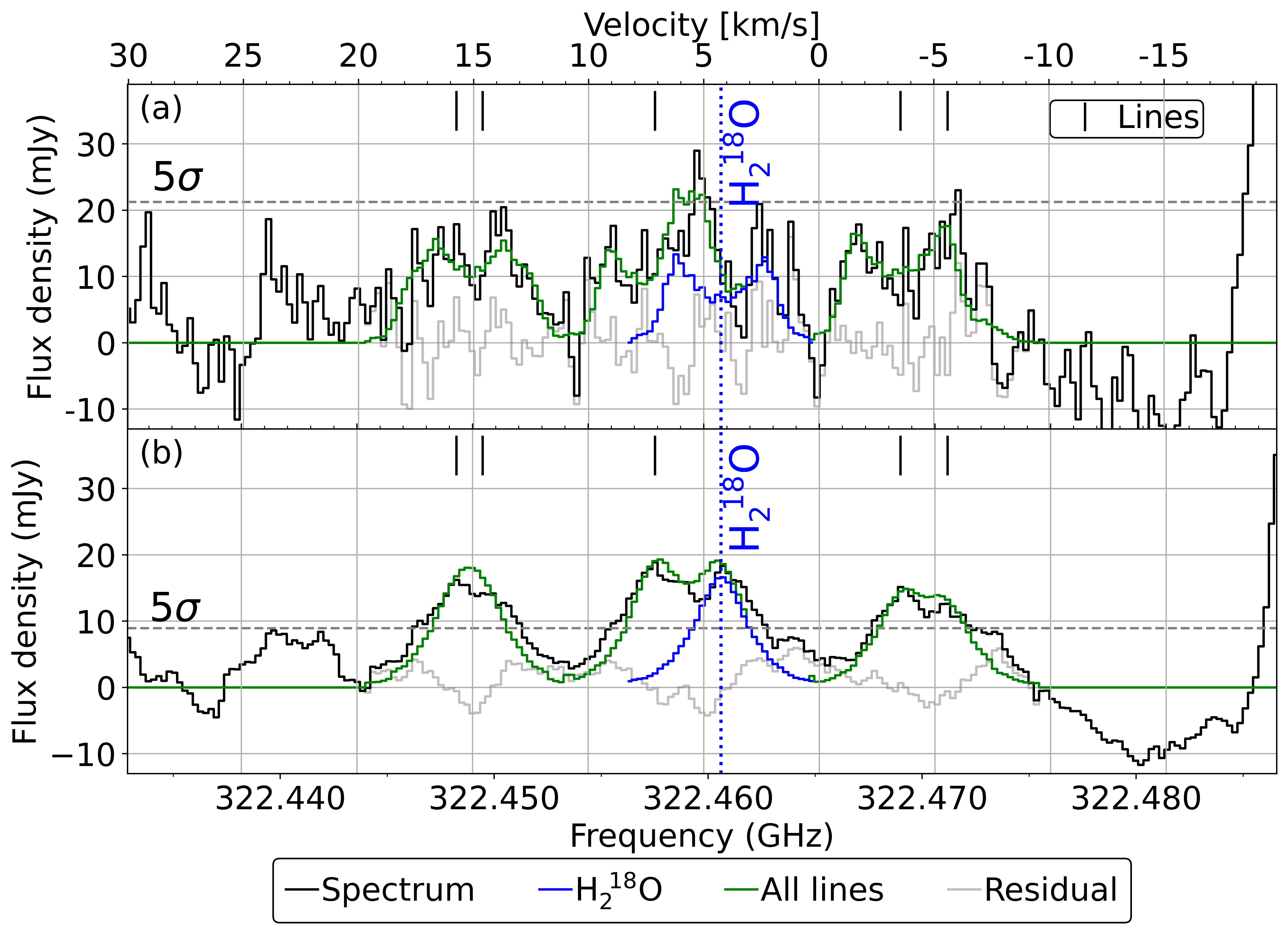}
\caption{\label{fig:spectra_h218o}
Similar to Figure~\ref{fig:spectra_h216o}, but for the H$_2$$^{18}$O 322 GHz line.
The blue solid lines show the fitted models for the H$_{2}$$^{18}$O line.
The green solid lines represent the combined fitting models of the H$_{2}$$^{18}$O line and the nearby contaminated lines.
The vertical black ticks indicate the central frequencies of the nearby contaminated lines (see Line 7--11 in Table~\ref{tab:kep_mask}).}
\end{figure*}

\begin{figure*}[htb!]
\centering
\includegraphics[width=0.7\textwidth]{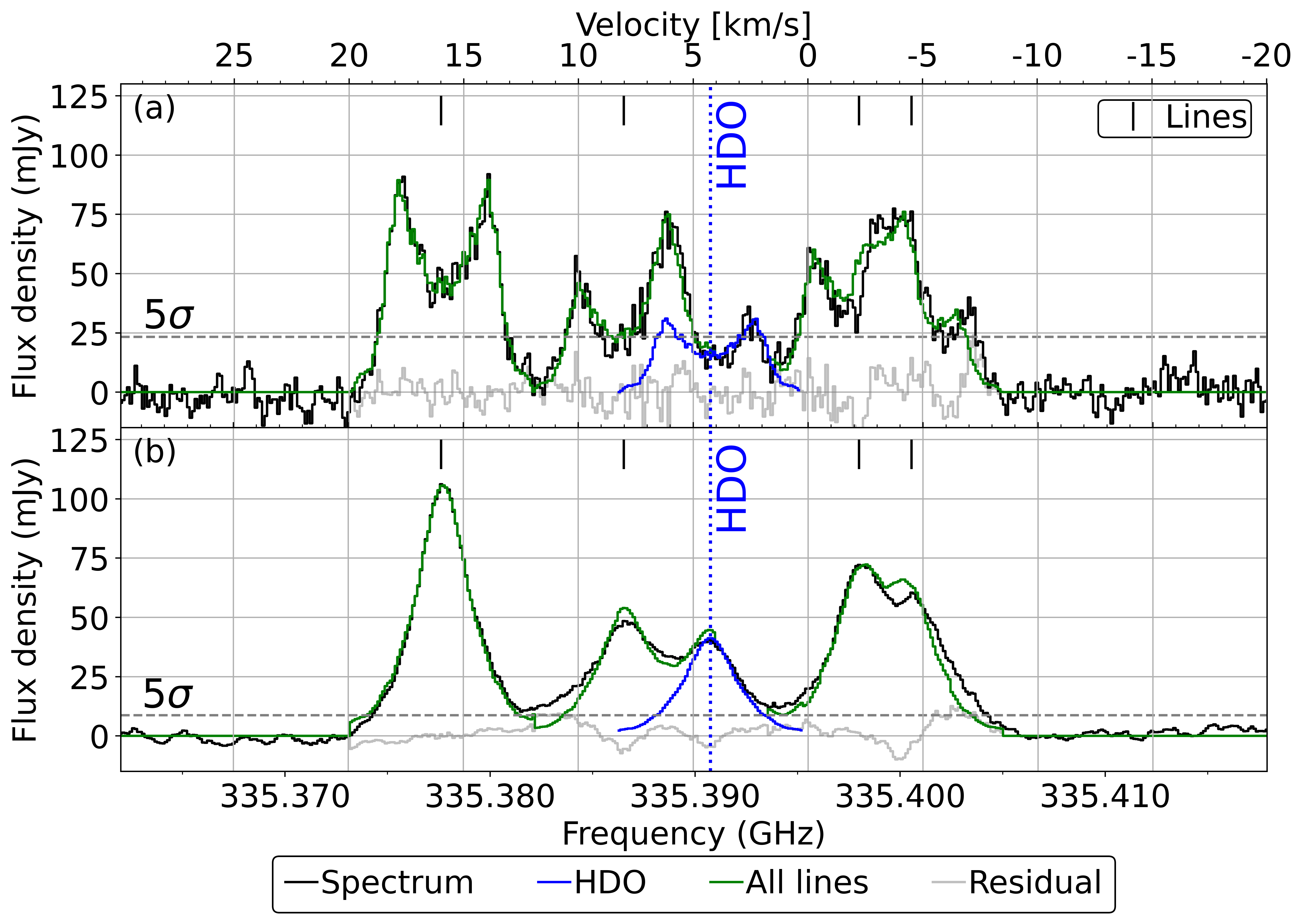}
\caption{\label{fig:spectra_hdo}
Similar to Figure~\ref{fig:spectra_h218o}, but for the HDO 335 GHz line.
The vertical black ticks indicate the central frequencies of the nearby contaminated lines (see Line 12--14 in Table~\ref{tab:kep_mask}).}
\end{figure*}

The resulting data cubes were used to generate spectra by integrating over an aperture with a radius of 0\farcs4.
Figures~\ref{fig:spectra_h216o}, \ref{fig:spectra_h218o}, and~\ref{fig:spectra_hdo} show the spectra obtained for detecting the H$_2$$^{16}$O, H$_2$$^{18}$O, and HDO lines, respectively.
In the top panels of Figures~\ref{fig:spectra_h216o} \ref{fig:spectra_h218o}, and~\ref{fig:spectra_hdo}, double-peaked line profiles are visible, reflecting the Keplerian rotation of the disk.

Next, to increase the sensitivity of the line detection and to resolve blended lines, we applied a correction for Keplerian rotation.
The correction assumes that the disk gas is in Keplerian rotation and removes the Doppler shifts associated with this orbital motion.
The spectra were extracted from a radius of 0\farcs4 using the \texttt{integrated\_spectrum()} function from the \texttt{GoFish} Python package \citep{2019_teague}, which deprojects the disk and shifts each line to the systemic velocity (4.25~km~s$^{-1}$), producing single-peaked profiles and allowing for easier identification by reducing blended components.
To correct for Keplerian rotation, the stellar mass, distance, inclination and position angle (1.29~$M_\odot$, 400~pc, 38.3\textdegree\, and 32\textdegree\, respectively) are provided as input parameters to the $\texttt{integrated\_spectrum()}$ function.
These values allow the function to deproject the disk and apply velocity alignment and line broadening correction accordingly.
In the $\texttt{integrated\_spectrum()}$ function, an inclination of 38.3\textdegree\ and a position angle of 32\textdegree\ are assumed, based on previous studies \citep{2016_cieza, 2023_tobin}.
We further assume that the emission arises from the disk midplane and neglect any vertical structure.
The resulting spectra are shown in the bottom panels of Figures~\ref{fig:spectra_h216o}, \ref{fig:spectra_h218o}, and~\ref{fig:spectra_hdo}.
After the correction for Keplerian rotation, each line exhibits a single-peaked profile, demonstrating that the Doppler shifts associated with the disk’s orbital motion have been properly removed.

The observed spectra and those corrected for Keplerian rotation, without the overlaid model components, are presented separately in Appendix Figures~\ref{fig:spectra_unstack} and~\ref{fig:spectra_stack}.

\subsection{Spectral Fitting and Line Identification}\label{Spectral Fitting and Line Identification}
In these single-peaked spectra, obtained by correcting for Keplerian rotation, we attributed spectral features to nearby contaminated lines if they exhibited peaks exceeding a 5$\sigma$ significance threshold and corresponded to transitions that had not yet been assigned.
To remove contamination from water lines caused by nearby contaminated lines, we constructed template spectra that could be applied to each detected line.

Specifically, to model the emission features of HDO, H$_2$$^{18}$O, and nearby contaminated lines, which could be COMs lines, we created a template spectrum using an isolated CH$_3$CHO vt=0 18$_{0,18}$-17$_{0,17}$, E 335.3825~GHz line.
This line had been previously identified and assigned in the same spectral window by \citet{2019_lee} as part of their line survey.

After subtracting the baseline to remove any remaining slope, the spectrum was scaled so that its peak intensity equals 1.0.
To generate a symmetric template, the spectrum was mirrored about its center of the line and averaged with the original profile, producing an idealized spectral shape suitable for modeling.
We constructed template spectra using both the double-peaked profile obtained before correcting for Keplerian rotation and the single-peaked profile obtained after the correction.

To accurately identify the water isotopologue lines, it was necessary to first remove the contribution from nearby contaminated lines that blend with them.
To this end, we began by determining the central frequencies of the blended lines using the Keplerian-corrected spectra, which provide deblended, single-peaked profiles suitable for line identification.
In Figures~\ref{fig:spectra_h216o}, \ref{fig:spectra_h218o} and ~\ref{fig:spectra_hdo}, these $>5\sigma$ peaks are marked by short black ticks above the spectra, and were treated as nearby contaminated lines.
A list of these nearby contaminated lines, together with their rest frequencies, is provided in Table~\ref{tab:kep_mask}.
Among the lines in Table~\ref{tab:kep_mask}, those in SPW~1 have central frequencies consistent with the transitions of $^{13}$CH$_3$OH and CH$_3$OCHO detected in Orion KL with ALMA Band 7 observations by \citet{2014_hirotaa} (see their Fig.~8), suggesting that these identifications are plausible.

\begin{deluxetable}{lc}
\tablewidth{0pt}
\tablecaption{Keplerian mask parameters for water isotopologues and contaminated lines blended with water\label{tab:kep_mask}}
\tablehead{
\colhead{Line} & 
\colhead{Rest freq.}
\\
\colhead{} & 
\colhead{(GHz)}
}
\startdata
\hline
\textbf{SPW 1}
\\
\hline
Line 1&
321.2028
\\
Line 2&
321.2152
\\
Line 3&
321.2233
\\
(H$_{2}$$^{16}$O 10$_{2,9}$-9$_{3,6}$)\tablenotemark{a} &
(321.2257)
\\
Line 4&
321.2291 
\\
Line 5&
321.2350
\\
Line 6&
321.2434
\\
\hline
\multicolumn{2}{l}{\textbf{SPW 2}}
\\
\hline
Line 7&
322.4528
\\
Line 8&
322.4540
\\
Line 9&
322.4621
\\
H$_{2}$$^{18}$O 5$_{1,5}$-4$_{2,2}$&
322.4652
\\
Line 10&
322.4736
\\
Line 11&
322.4758
\\
\hline
\multicolumn{2}{l}{\textbf{SPW 3}}
\\
\hline
CH$_3$CHO vt=0, 18$_{0, 18}$-17$_{0, 17}$, E&
335.3825
\\
Line 12&
335.3913
\\
HDO 3$_{3,1}$-4$_{2,2}$&
335.3955
\\
Line 13&
335.4028
\\
Line 14&
335.4053
\\
\enddata
\tablenotetext{a}{H$_2$$^{16}$O was not detected.}
\tablecomments{
SPWs 1, 2, and 3 correspond to the spectral windows targeting the H$_2$$^{16}$O 321 GHz, H$_2$$^{18}$O 322 GHz, and HDO 335 GHz lines, respectively.
For all lines, the Keplerian mask was generated using the \texttt{keplerian\_mask} code \citep{2020_teague}, adopting the following parameters: Smoothing = $0.65$, $\text{R}_{\text{max}} = 0\farcs3$, $\text{R}_{\text{min}} = 0\farcs1$, $\text{Z}/\text{R} = 0.4$, and $\Delta \text{V} = 0.1~\text{km s}^{-1}$.
Line 1-14 denotes an unidentified spectral feature associated with a 5$\sigma$ peak in the stacked spectrum.
}
\end{deluxetable}

The spectral modeling was performed by fitting the observed spectra as a linear combination of the scaled template profiles.
Using the identified frequencies of each component as the above method, we employed the \texttt{scipy.optimize.minimize} function to determine the best-fitting multiplicative scaling factors.
The results of this fitting procedure are presented in Figures~\ref{fig:spectra_h216o}, \ref{fig:spectra_h218o} and~\ref{fig:spectra_hdo}.
In these figures, the green curves represent the best-fit models including both nearby contaminated lines and water emission components, while the blue curves isolate the contribution from the water lines alone.
Using this approach, HDO was detected at 6.7$\sigma$ and 23.6$\sigma$ before and after correction for Keplerian rotation, respectively.
H$_2$$^{18}$O was tentatively detected at 3.1$\sigma$ before the correction, and successfully detected at 9.3$\sigma$ after the correction.
In contrast, H$_2$$^{16}$O was not detected.
Although it should be more abundant than the other two isotopologues, its 321 GHz transition has a much higher upper state energy than those in other two observed H$_{2}$$^{18}$O and HDO transitions (see Table~\ref{tab:parameters}); as a result, its emission is likely buried beneath the upper limits of other adjacent transition lines.

Following the spectral fitting, we derived the integrated intensities of the HDO and H$_{2}$$^{18}$O lines by integrating the model spectra corresponding to the water line components.
Specifically, the integration was performed over the blue curves shown in Figures~\ref{fig:spectra_h218o} and~\ref{fig:spectra_hdo}.
For the o-H$_2$$^{16}$O 321 GHz line, which was not detected in this study, we estimated the integrated intensity that the modeled double-peaked spectrum would yield if its peaks reached the 3$\sigma$ level, and adopted this value as the upper limit.
The results are summarized in Table~\ref{tab:line_flux}.

We note that our approach differs from that of \citet{2023_tobin}, who used the isolated CH$_3$OH 242 GHz lines detected in Band 6 to construct a template spectrum, and derived the integrated intensity by subtracting the modeled COM emission from the observed spectrum and integrating the residuals.
We performed the fitting using template spectra from both CH$_3$CHO 335 GHz and CH$_3$OH 242 GHz.
Nevertheless, the integrated intensities of the fitted water model spectra are consistent within 1$\sigma$ uncertainties (see Appendix~\ref{app:appendixB}).

\begin{deluxetable}{lcll}
\tablewidth{0pt}
\tablecaption{Integrated intensity, $T_\mathrm{rot}$, and $N_\mathrm{thin}$\label{tab:line_flux}}
\tablehead{
\colhead{Isotope} &
\colhead{$I_\mathrm{int}$} &
\colhead{$T_\mathrm{rot}$} &
\colhead{$\log_{10}(N_{\mathrm{thin}})$} 
\\
\colhead{} &
\colhead{($\mathrm{mJy\,km\,s^{-1}}$)} &
\colhead{(K)} &
\colhead{(cm$^{-2}$)} 
}
\startdata
\hline
(o-H$_2$$^{16}$O)\tablenotemark{a} &
$\leq60.89$
\\
\hline
p-H$_2$$^{18}$O &
$52.06\pm7.48$\\
\hline
\multicolumn{2}{r}{Median:}&
$119.33^{+15.57}_{-11.98}$ &
$15.68^{+0.09}_{-0.09}$
\\
\multicolumn{2}{r}{MAP:}&
$116.89\pm12.81$ &
$15.69\pm0.15$
\\
\hline
HDO &
$123.73\pm6.15$
\\
\hline
\multicolumn{2}{r}{Median:} &
$87.93^{+5.23}_{-4.78}$ &
$15.65^{+0.03}_{-0.03}$
\\
\multicolumn{2}{r}{MAP:} &
$87.46\pm4.95$ &
$15.65\pm0.06$
\\
\enddata
\tablenotetext{a}{3$\sigma$ uppper limits are estimated for non-detection.}
\end{deluxetable}

\begin{figure*}[htb!]
\centering
\includegraphics[width=0.7\textwidth]{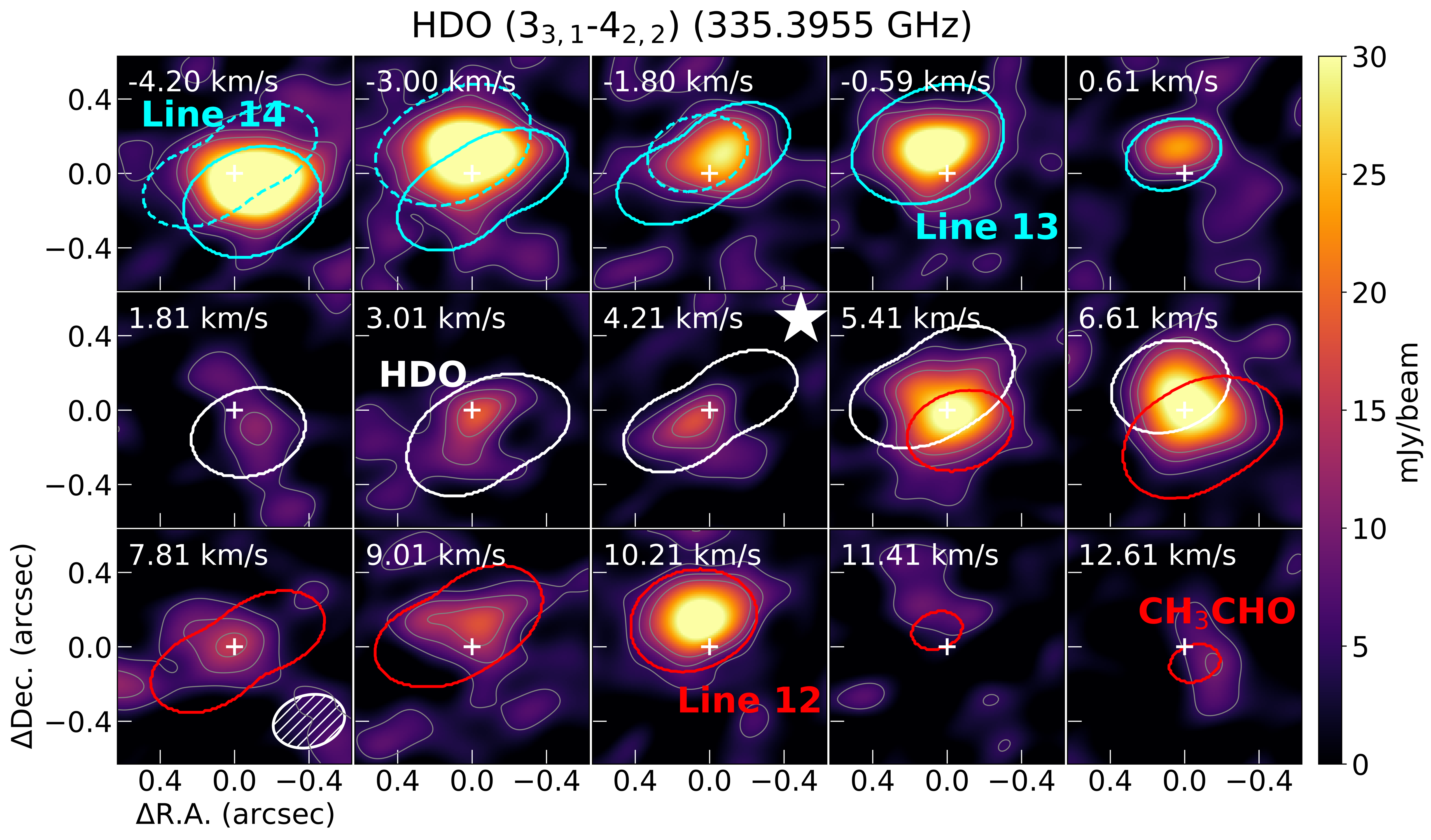}
\caption{\label{fig:channel_hdo}
Channel maps around the frequency of the HDO line at 335 GHz.
The position of the protostar is marked with a white cross, and the channels closest to the systemic velocity of V883 Ori (4.25 km s$^{-1}$) are indicated by a star in the upper right corner.
Thin gray contours represent 1, 2, and 3$\sigma$ levels.
To clearly highlight the emission region of the HDO line, the emission is saturated in some panels.
The synthesized beam is shown in the lower right corner of the bottom-left panel.
The Keplerian mask for the HDO emission is drawn as a thick white line, and the masks corresponding to the blended nearby contaminated lines and COM line; two in blueshifted region and two in redshifted region of the HDO line are shown as thick cyan and red lines, respectively.
}
\end{figure*}

\begin{figure*}[htb!]
\centering
\includegraphics[width=0.7\textwidth]{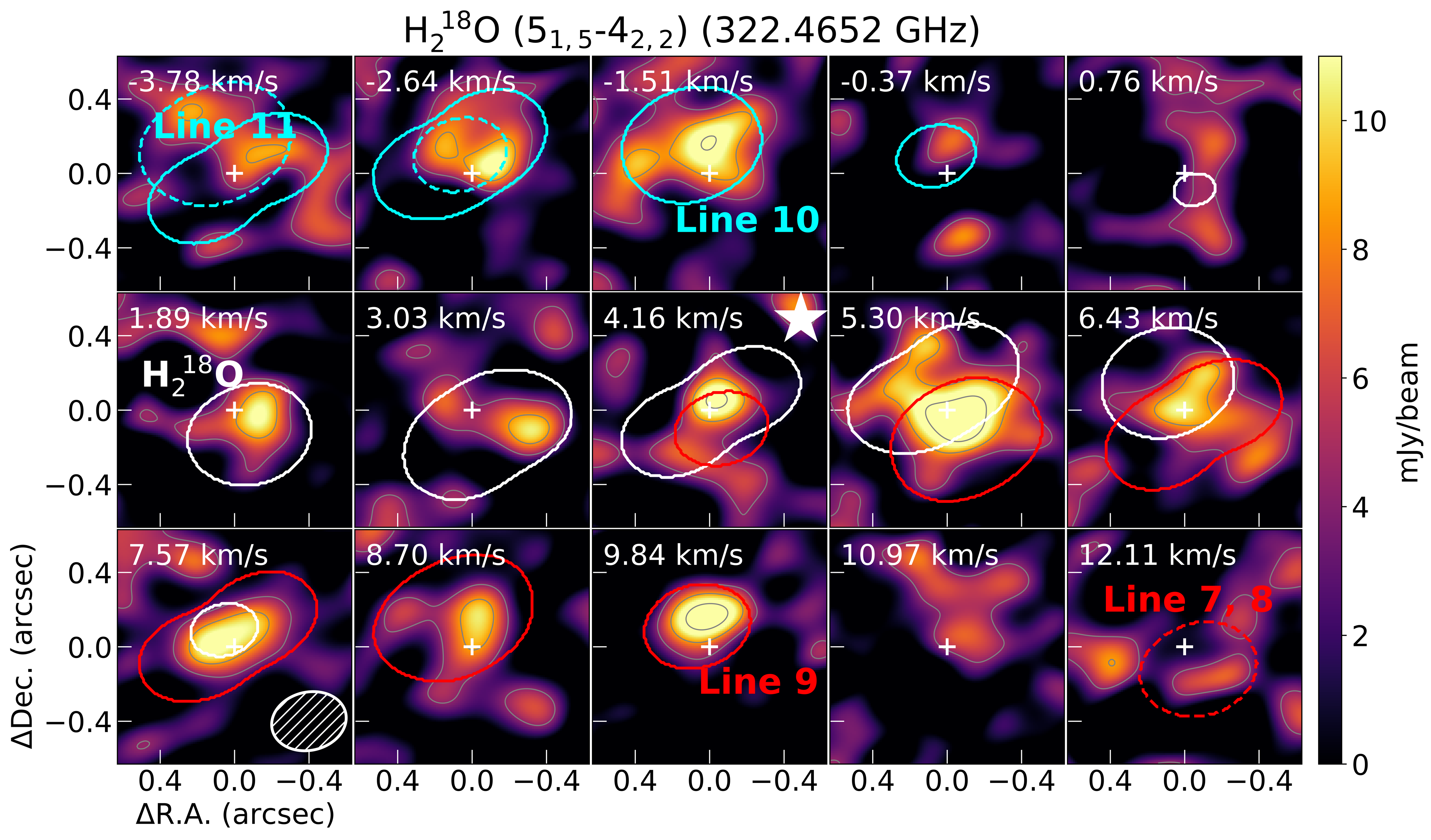}
\caption{\label{fig:channel_h218o}
Similar to Figure~\ref{fig:channel_h218o}, but for the H$_{2}$$^{18}$O line at 322 GHz.
The Keplerian mask for the H$_{2}$$^{18}$O emission is drawn as a thick white line, and the masks corresponding to the nearby contaminated lines; two in blueshifted region and three in redshifted region of the H$_{2}$$^{18}$ are shown as thick cyan and red lines, respectively.
}
\end{figure*}

\subsection{Spatial Distributions}\label{Spatial Distributions}
\subsubsection{Channel Maps}\label{Channel Maps}
Figures~\ref{fig:channel_hdo} and~\ref{fig:channel_h218o} present the channel maps of the HDO and H$_{2}$$^{18}$O emission, overlaid with the corresponding Keplerian masks generated with the \texttt{keplerian\_mask} code \citep{2020_teague}.
Keplerian masks for the contaminating lines are also shown to indicate the expected velocity ranges and spatial locations of overlapping emission.
The parameters used to construct these masks are listed in Table~\ref{tab:kep_mask}, and were applied consistently across both the water isotopologue lines and the nearby contaminated lines.
As evident in the spectra presented in Figures~\ref{fig:spectra_h216o}, \ref{fig:spectra_h218o}, and~\ref{fig:spectra_hdo}, the target water lines exhibit partial blending with nearby contaminated lines, which is also apparent in the channel maps.
However, for both HDO and H$_{2}$$^{18}$O, minimal contamination is seen within the velocity range 1.45–3.45 km s$^{-1}$.
Within this range, the emission is expected to originate predominantly from the water lines, allowing for a robust extraction of uncontaminated water emission.

\subsubsection{Moment 0 Extraction}\label{Moment 0 Extraction}
\begin{figure*}[htb!]
\centering
\includegraphics[width=0.7\textwidth]{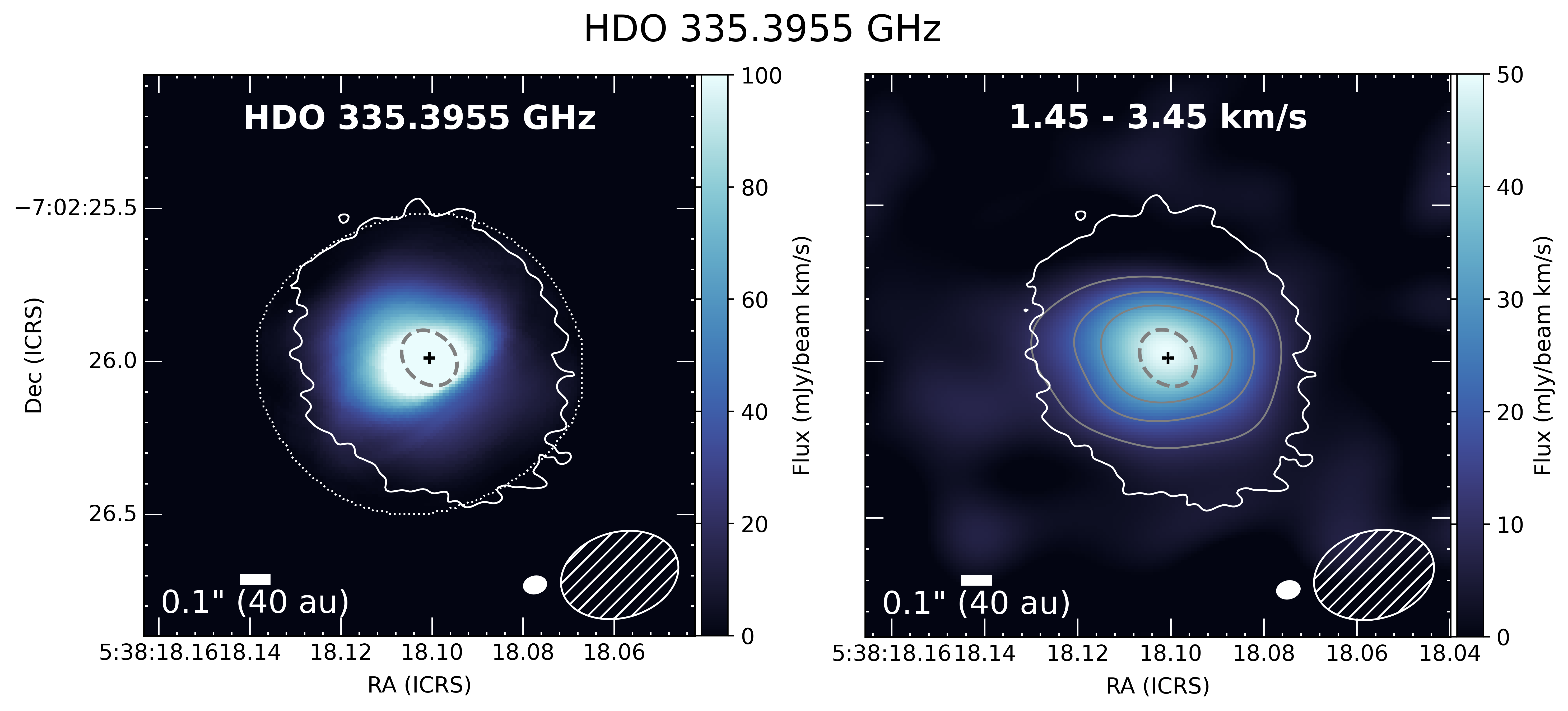}
\caption{\label{fig:moment0_hdo} 
The HDO 335 GHz integrated intensity map created using a Keplerian mask (the left panel) and that integrated over the velocity range 1.45--3.45 km s$^{-1}$ (the right panel). 
The contours in the right panel represent intensity levels at 5$\sigma$, 10$\sigma$ and 15$\sigma$, where 1$\sigma$ corresponds to 2.01 mJy beam$^{-1}$ km s$^{-1}$, computed by integrating over 1.45--3.45 km s$^{-1}$.
The 1.45--3.45 km s$^{-1}$ velocity range was selected to minimize contamination from other lines.
The position of the protostar is marked with a black cross.
The optically thick region suggested by previous studies is indicated by the thick gray line at the center of each panel.
The hatched ellipses in the lower right corners denote the synthesized beam size of the line observation ($\sim 0\farcs 28 \times 0\farcs 39$)
The solid white line outlines the emission region of the dust continuum, while the filled white ellipse indicates the synthesized beam of the dust continuum observation ($\sim0\farcs08$).
In the left panel, the dotted white line marks the outer extent of the Keplerian mask.}
\end{figure*}

\begin{figure*}[htb!]
\centering
\includegraphics[width=0.7\textwidth]{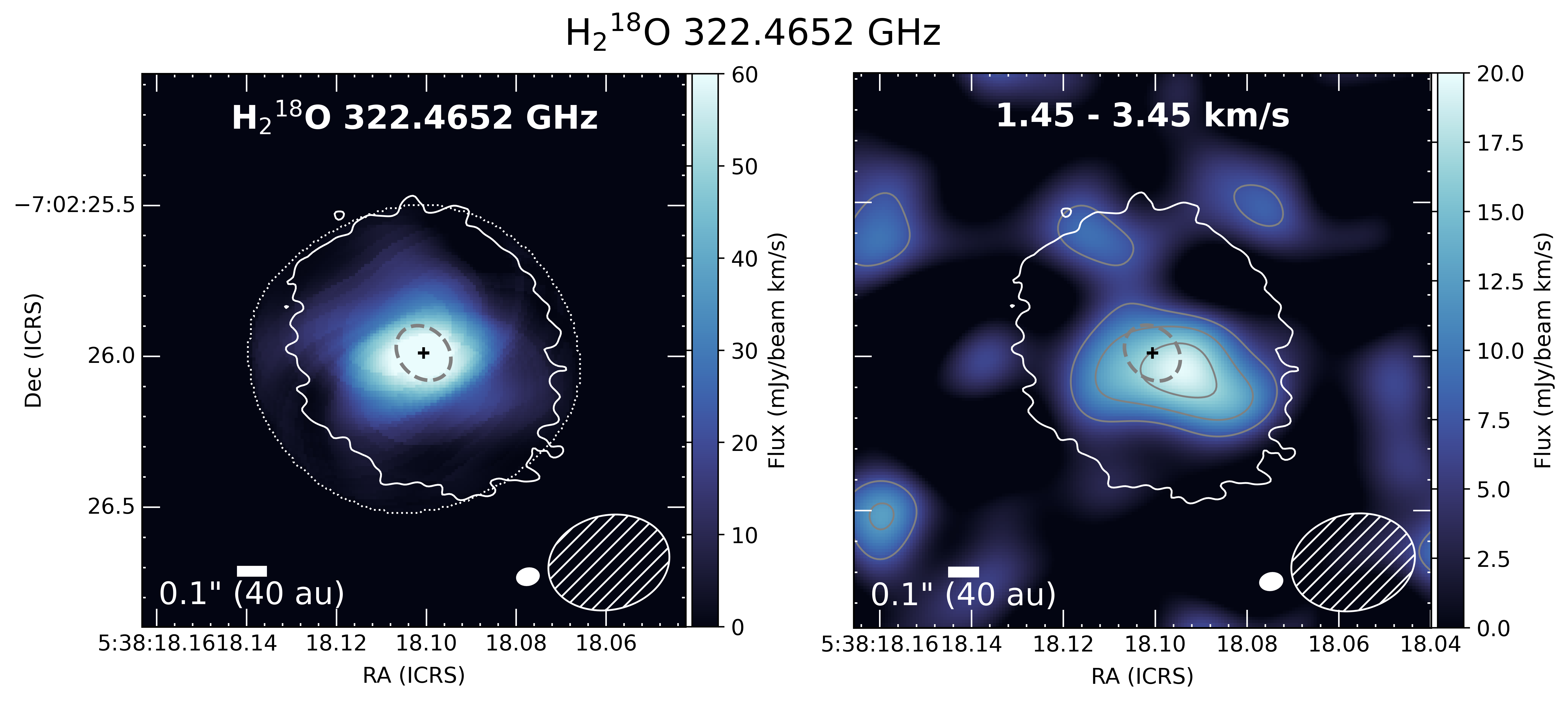}
\caption{\label{fig:moment0_h218o}
Similar to Figure~\ref{fig:moment0_hdo}, but for H$_{2}$$^{18}$O 322 GHz line.
The contours in the right panel represent intensity levels at 3$\sigma$, 5$\sigma$ and 7$\sigma$, where 1$\sigma$ corresponds to 2.39 mJy beam$^{-1}$ km s$^{-1}$, computed by integrating over 1.45--3.45 km s$^{-1}$.
The hatched ellipses in the lower right corners denote the synthesized beam size of the line observation ($\sim0\farcs 31 \times 0\farcs 40$).
}
\end{figure*}

Figures~\ref{fig:moment0_hdo}  and~\ref{fig:moment0_h218o} show the velocity-integrated intensity maps (moment 0 maps) of the selected water isotopologue transitions.
In each figure, the left panel presents the moment 0 map generated using the Keplerian mask defined by the parameters listed in Table \ref{tab:kep_mask} for the target water line.
The right panel shows the moment 0 map integrated over a velocity range (1.45–-3.45 km s$^{-1}$) identified from the channel maps as being minimally contaminated by other lines; no masking was applied except for this velocity selection.
The spatial distribution of the water line emission indicates that it originates from within the disk region (approximately 0\farcs3 or $\sim$120 au in radius).
However, the emission is only marginally spatially resolved at the current resolution of $\sim0\farcs3$–-$0\farcs4$, and the central cavities reported in the Band 5 and Band 7 data by \citet{2023_tobin} are not detected due to the limitation of the lower spatial resolution.

\subsection{Rotation Temperature and Column Density}\label{Rotation Temperature and Column Density}
\begin{figure*}[htb!]
\centering
\begin{minipage}[b]{0.48\textwidth}
  \centering
  \includegraphics[width=0.8\textwidth]{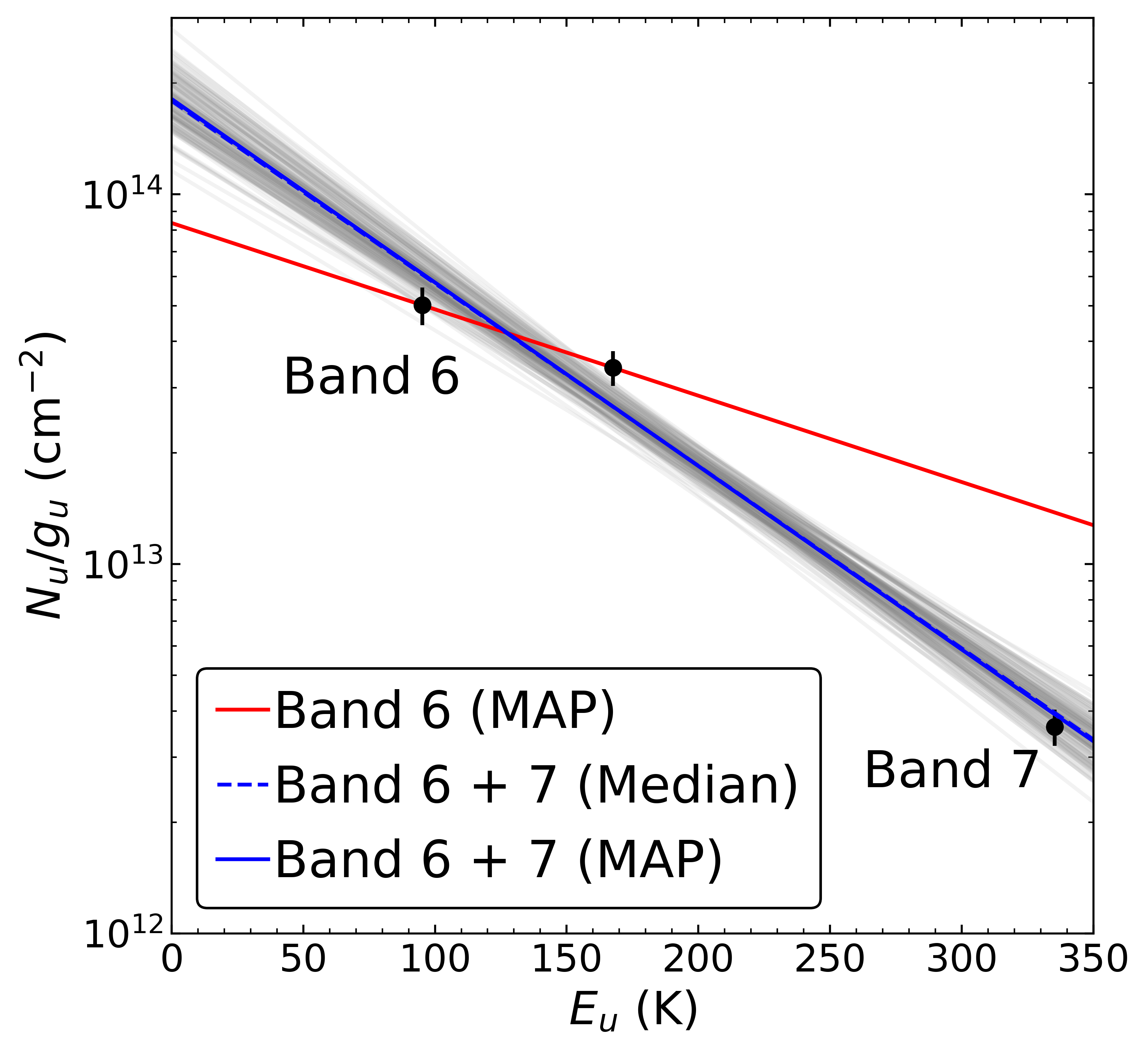}
\end{minipage}
\hfill
\begin{minipage}[b]{0.48\textwidth}
  \centering
  \includegraphics[width=0.8\textwidth]{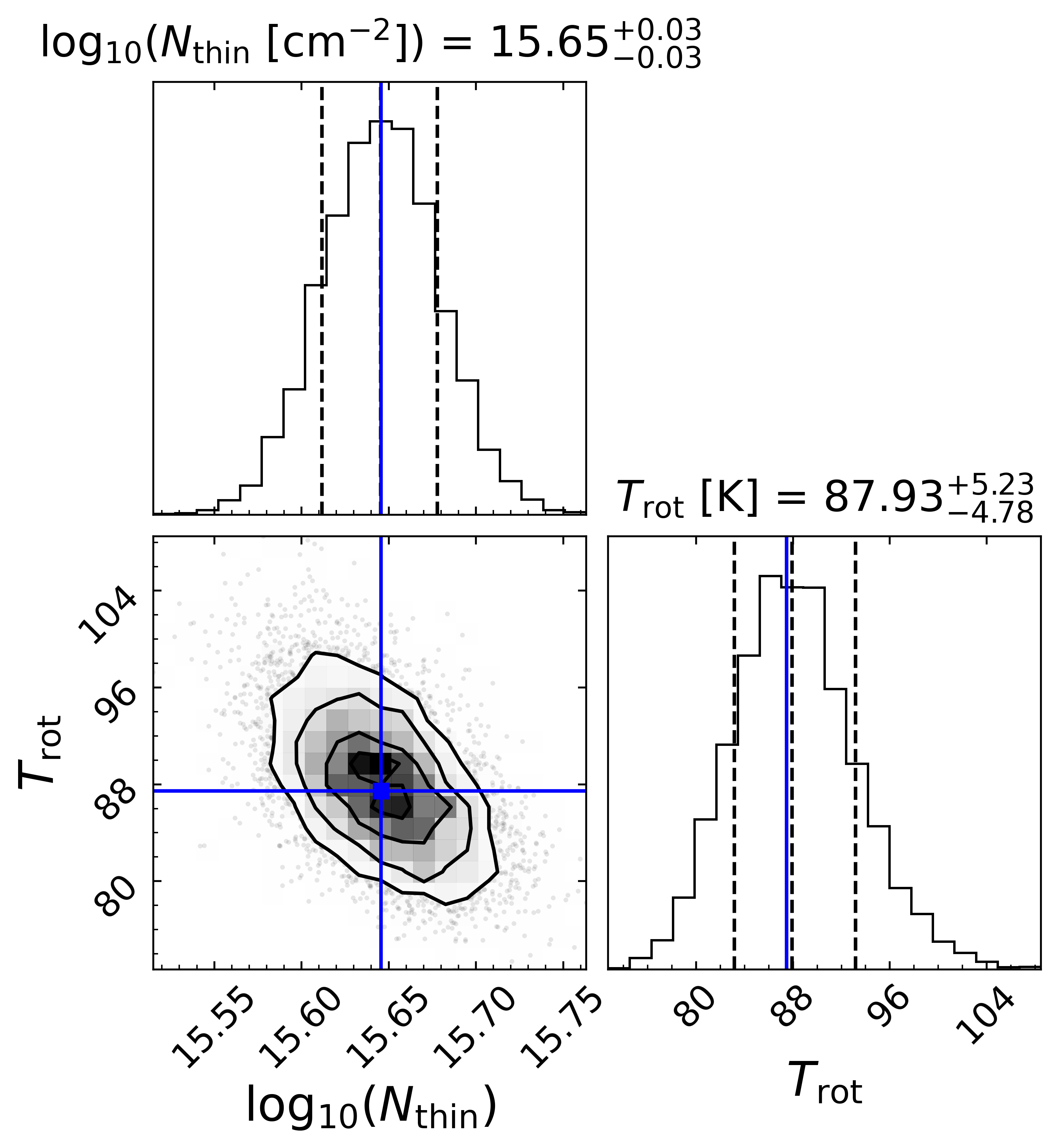}
\end{minipage}
\caption{\label{fig:rotation_hdo}%
Left: Rotational diagram based on three HDO lines observed with ALMA Bands 6 and 7. The blue dashed line represents the median result from the MCMC sampling, while the blue solid line shows the best-fit model corresponding to the maximum a posteriori (MAP) estimate; the dashed and solid lines almost overlap. Gray solid lines indicate a subset of MCMC samples, illustrating the uncertainty in the fit. For comparison, the red solid line shows the MAP estimate derived from an MCMC analysis using only the two Band 6 HDO lines reported by \citet{2023_tobin} (see Appendix~\ref{app:appendixC} for the rotational diagram).
Right: Corner plot showing the results of the MCMC sampling for the rotational diagram fitting based on three HDO lines. The black dashed lines indicate the 16th, 50th, and 84th percentiles of the posterior distributions, with the corresponding values shown at the top of each panel, while the blue solid lines represent MAP estimates.
}
\end{figure*}

\begin{figure*}[htb!]
\centering
\begin{minipage}[b]{0.48\textwidth}
  \centering
  \includegraphics[width=0.8\textwidth]{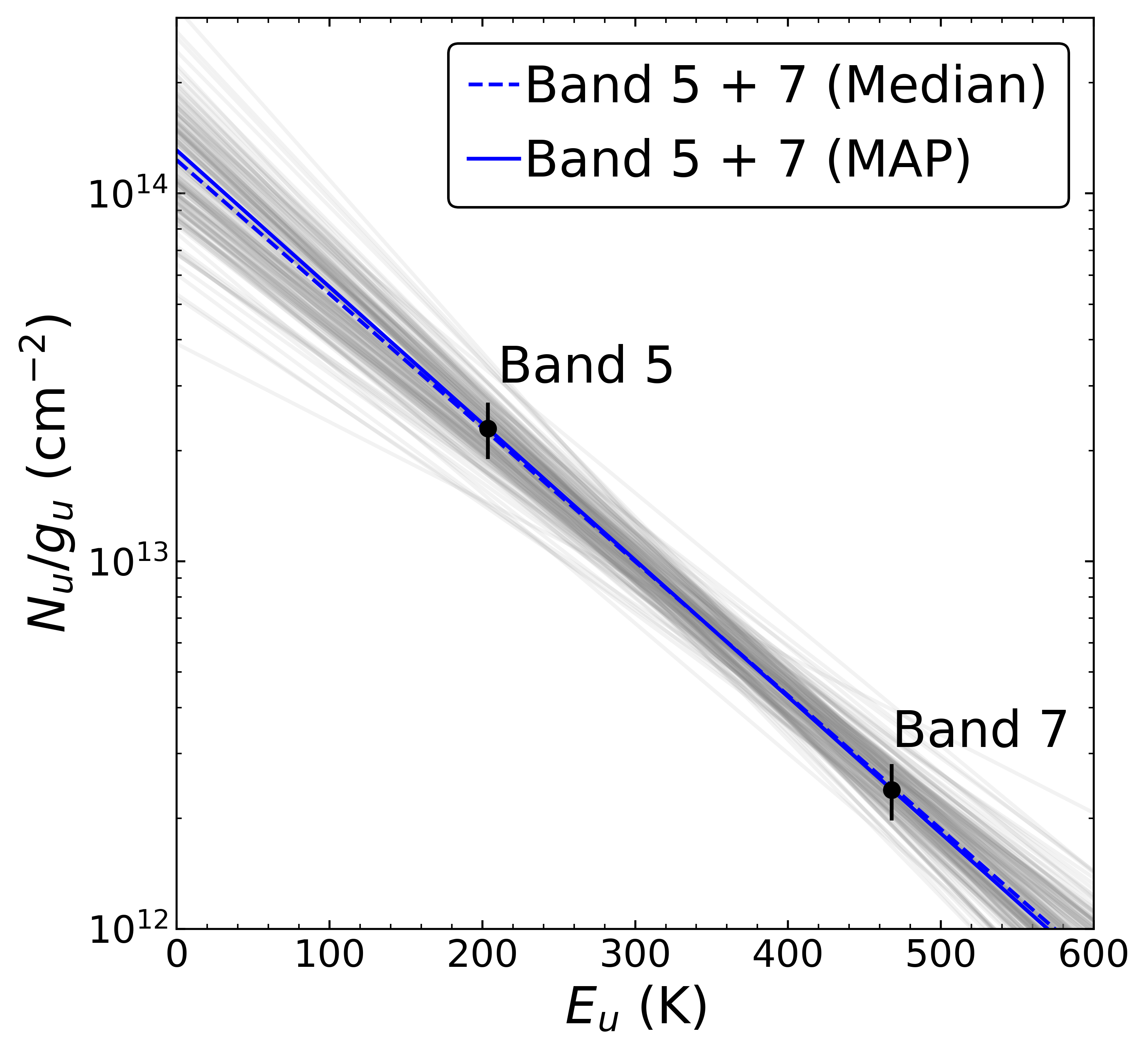}
\end{minipage}
\hfill
\begin{minipage}[b]{0.48\textwidth}
  \centering
  \includegraphics[width=0.8\textwidth]{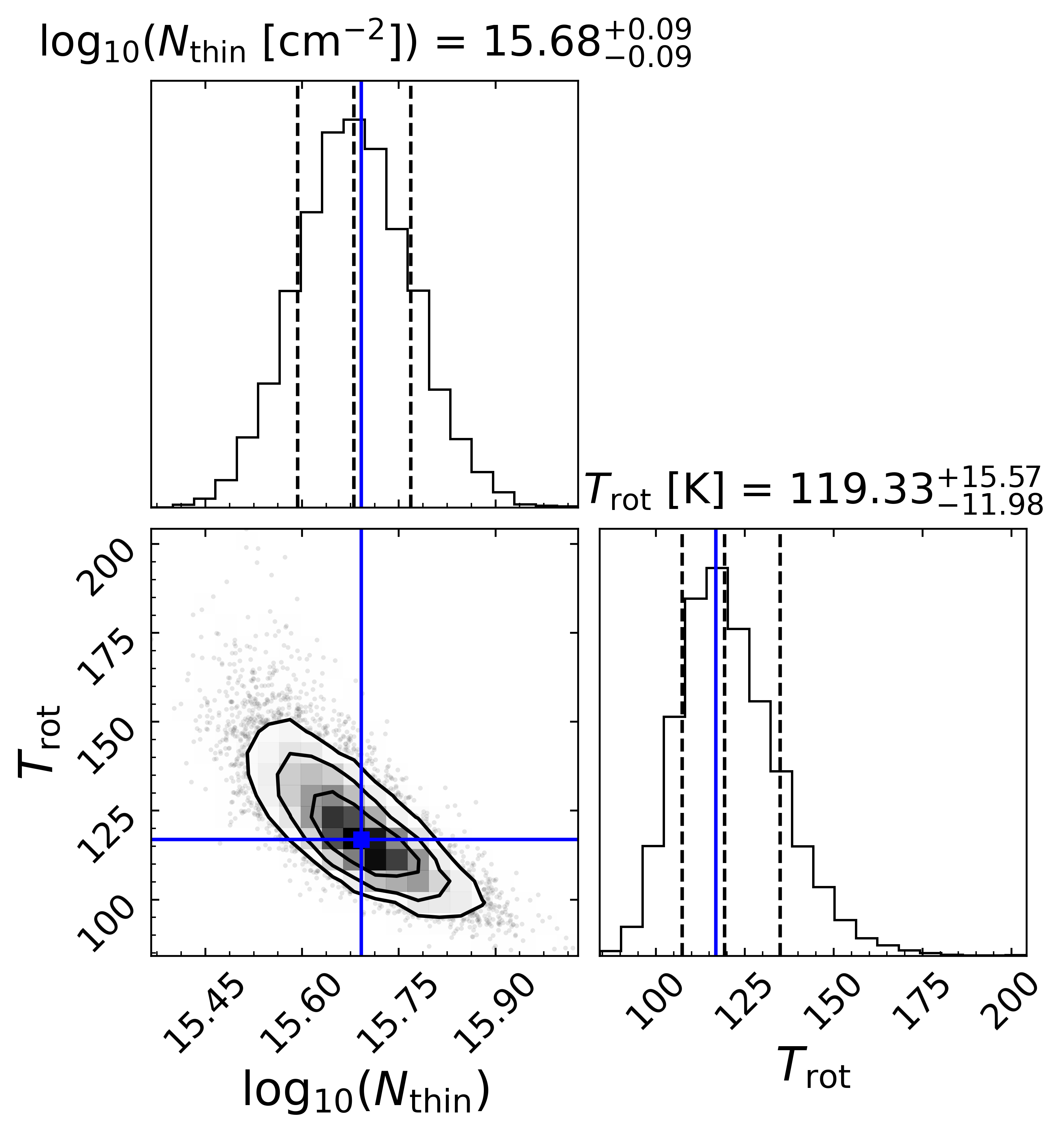}
\end{minipage}
\caption{\label{fig:rotation_h218o}%
The same as Figure~\ref{fig:rotation_hdo}, but for two H$_{2}$$^{18}$O lines observed with ALMA Bands 5 and 7.}
\end{figure*}

In Figures~\ref{fig:rotation_hdo} and~\ref{fig:rotation_h218o}, we constructed rotational diagrams for HDO and H$_{2}$$^{18}$O by combining our new ALMA Band 7 observations of the HDO 335 GHz and H$_{2}$$^{18}$O 322 GHz lines with the HDO 225 and 241 GHz transitions in Band 6 and the H$_{2}$$^{18}$O 203 GHz line in Band 5 reported by \citet{2023_tobin}.

We adopted the assumptions of optically thin emission and a uniform excitation temperature across the relevant energy range.
Our methodology follows that of \citet{2024_facchini}.

Partition functions as a function of temperature were taken from the JPL database \citep{1998_pickett} for HDO and H$_2$$^{18}$O, interpolating between tabulated values when necessary.
We employed the full partition function, which assumes an ortho-to-para ratio of 3:1 for H$_2$$^{18}$O, and therefore did not rescale individual ortho and para line fluxes.

We performed a linear regression in the $\log(N/g_\mathrm{u})$ versus $E_\mathrm{u}$ space using the \texttt{emcee} sampler to derive the rotational temperature $T_\mathrm{rot}$ and the column density $N$.
In this analysis, the column density in each energy level $N_\mathrm{u}$ is related to the total column density $N$ through
\begin{equation}\label{eq:1}
N
=
\frac
{N_\mathrm{u}}
{g_\mathrm{u}}
Q_\mathrm{rot}(T_\mathrm{ex})
e^{\frac{E_\mathrm{u}}{T_\mathrm{ex}}}
\end{equation}
where $g_\mathrm{u}$ is the statistical weight of the upper state, $E_\mathrm{u}$ is its energy, and $Q_\mathrm{rot}(T_\mathrm{ex})$ is the rotational partition function at the excitation temperature $T_\mathrm{ex}$.
By fitting a straight line to $\log(N_\mathrm{u}/g_\mathrm{u})$ as a function of $E_\mathrm{u}$, the slope yields $1/T_\mathrm{rot}$, while the intercept provides $N$.
This approach assumes local thermodynamic equilibrium (LTE) and that the lines are optically thin.
The sampling employed 32 walkers, 5,000 steps, and a 3,000-step burn-in. Molecular coefficients were taken from Table~\ref{tab:parameters}.
In the optically thin assumption, $N_\mathrm{thin}$ represents the column density and $T_\mathrm{rot}$ the rotational temperature, where $E_\mathrm{u}$ is the upper state energy and $g_\mathrm{u}$ the upper state degeneracy.
Each rotational diagram was fitted with best-fit parameters for $N_\mathrm{thin}$ and $T_\mathrm{rot}$ for HDO and H$_{2}$$^{18}$O, respectively.

The marginalized posterior distributions obtained from the MCMC sampling are shown in Figures~\ref{fig:rotation_hdo} and~\ref{fig:rotation_h218o}, with the corresponding corner plots displayed in the right panels of each figure.
For each molecule, both the median values (with uncertainties corresponding to the 16th and 84th percentiles) and the maximum a posteriori (MAP) estimates are reported.
The results of the MAP estimation indicate that H$_2$$^{18}$O has a rotational temperature of $116.89\pm12.81$ K and a logarithmic column density of $\log(N [\mathrm{cm}^{-2}]) = 15.69\pm0.15$, whereas HDO shows a lower rotational temperature of $87.46\pm4.95$ K and a logarithmic column density of $\log(N [\mathrm{cm}^{-2}]) = 15.65\pm0.06$.
Based on these values, the optical depths of the lines are on the order of $2\times10^{-2}$, consistent with the assumption that they are optically thin.
The derived excitation temperatures and column densities, along with the integrated line intensities, are summarized in Table~\ref{tab:line_flux}.
All uncertainties include the 10 \% absolute flux calibration uncertainty of ALMA.

In the rotational diagram for HDO, the 335 GHz line observed in ALMA Band 7 falls significantly below the linear trend defined by the two transitions detected in Band 6 (at 225 and 241 GHz).
This deviation indicates that the Band 7 line is substantially weaker than expected based on the rotational temperature and column density derived from the Band 6 data.
Quantitatively, the observed intensity of the Band 7 transition is only 26\% of the value predicted by extrapolating the linear fit from the two Band 6 lines to the upper state energy of the 335 GHz transition.
In other words, the line is roughly a quarter as strong as expected.
As a result, including the Band 7 line in the fit yields a significantly lower rotational temperature of $\sim$87.4 K, compared to 186.4 K derived from the Band 6 data alone (see Figure~\ref{fig:rotation_hdo_band6}).
For the rotational diagram constructed from the two H$_2$$^{18}$O lines observed in Bands 5 and 7, the derived excitation temperature is $T_\mathrm{ex} = 119$ K, lower than the $T_\mathrm{ex} = 203$ K inferred from the rotational diagram based on the two Band 6 HDO lines.
This suggests that the intensity of the Band 7 H$_2$$^{18}$O line may also be weaker than expected from the observations in Bands 5 and 6.
We interpret this discrepancy as a suppression of the Band 7 HDO line intensity in particular, and explore potential causes for this attenuation in Section~\ref{seq: Band 7 HDO Line Weakness Relative to Rotational Diagram Predictions}.

\subsection{Total Column Densities at Several Excitation Temperatures}\label{seq:total column density}
Based on the H$_{2}$$^{16}$O, H$_{2}$$^{18}$O and HDO lines in Bands 5, 6, and 7, the total column densities were calculated under the assumption of several excitation temperatures (75, 180, and 225 K) in order to investigate the water abundance and the D/H ratio under different thermal conditions.
The results are summarized in Table~\ref{tab:column_density}.
The derived total column density is significantly higher compared to those derived from other detected water lines especially when we assume $T_\mathrm{ex}=75$ K.
This is because the H$_2$$^{16}$O 321 GHz line has a much higher upper state energy compared to the other isotopic water lines (see Table~\ref{tab:parameters}).

Although the H$_2$$^{18}$O line intensities in Band 7 are weaker than those in Band 5 (see Table~\ref{tab:template}), the derived column density $N$ is larger in Band 7 at $T_\mathrm{ex}=75\ \mathrm{K}$.
This is because the upper state energy of the Band 5 H$_2$$^{18}$O line ($E_\mathrm{u}(203\ \mathrm{GHz}) = 203.7\ \mathrm{K}$) is much lower than that of the Band 7 H$_2$$^{18}$O line ($E_\mathrm{u}(322\ \mathrm{GHz}) = 467.9\ \mathrm{K}$).
When Equation~(\ref{eq:1}) is treated as a function of $T_\mathrm{ex}$, the exponential term gives a much larger contribution in Band 7, resulting in a higher value of the total column density $N$.
This trend is also observed in the comparison between the Band 6 and Band 7 HDO lines: the upper state energies of the Band 6 HDO lines are $E_\mathrm{u}(225\ \mathrm{GHz}) = 167.6\ \mathrm{K}$ and $E_\mathrm{u}(241\ \mathrm{GHz}) = 95.2\ \mathrm{K}$, whereas the Band 7 HDO line has $E_\mathrm{u}(335\ \mathrm{GHz}) = 335.3\ \mathrm{K}$, which is sufficiently higher.
As a result, the derived column density $N$ at $T_\mathrm{ex}=75\ \mathrm{K}$ is larger in Band 7.

Furthermore, for the Band 5 and Band 6 H$_2$$^{18}$O and HDO lines, the derived column density $N$ is larger at $T_\mathrm{ex}=225\ \mathrm{K}$ than at $T_\mathrm{ex}=180\ \mathrm{K}$.
In contrast, for the Band 7 H$_2$$^{18}$O and HDO lines, $N$ becomes smaller at $T_\mathrm{ex}=225\ \mathrm{K}$ than at $T_\mathrm{ex}=180\ \mathrm{K}$.
This behavior arises because the upper state energies $E_\mathrm{u}$ of the Band 7 lines are higher, and thus the decrease in the exponential term of Equation~(\ref{eq:1}) has a stronger impact.

\begin{deluxetable*}{lcccc}
\tablewidth{0pt}
\tablecaption{Total column density $N$ at different excitation temperatures $T_\mathrm{ex}$ \label{tab:column_density}}
\tablehead{
\colhead{Band} &
\colhead{Line} &
\multicolumn{3}{c}{Total column density (cm$^{-2}$)}
\\
&
&
\colhead{($T_\mathrm{ex} = 75\ \mathrm{K}$)} &
\colhead{($T_\mathrm{ex} = 180\ \mathrm{K}$)} &
\colhead{($T_\mathrm{ex} = 225\ \mathrm{K}$)}
}
\startdata
\textbf{Band 5} & H$_2$$^{18}$O (203 GHz) & $(8.11 \pm 1.42)\times 10^{15}$ & $(6.08 \pm 1.06)\times 10^{15}$ & $(6.65 \pm 1.16)\times 10^{15}$ \\
\hline
\textbf{Band 6} & HDO (225 GHz) & $(5.98\pm 0.64)\times 10^{15}$ & $(6.00\pm 0.64)\times 10^{15}$ & $(6.84\pm 0.73)\times 10^{15}$ \\
               & HDO (241 GHz) & $(3.36\pm 0.39)\times 10^{15}$ & $(5.92\pm 0.69)\times 10^{15}$ & $(7.31\pm 0.85)\times 10^{15}$ \\
\hline
\textbf{Band 7} & H$_2$$^{16}$O (321 GHz)\tablenotemark{a} & $\le 1.16\times 10^{24}$ & $\le 2.20\times 10^{18}$ & $\le 3.81\times 10^{17}$ \\
               & H$_2$$^{18}$O (322 GHz) & $(2.86\pm 0.50)\times 10^{16}$ & $(2.75\pm 0.48)\times 10^{15}$ & $(2.24\pm 0.39)\times 10^{15}$ \\
               & HDO (335 GHz) & $(5.98\pm 0.67)\times 10^{15}$ & $(1.63\pm 0.18)\times 10^{15}$ & $(1.54\pm 0.17)\times 10^{15}$ \\
\enddata
\tablenotetext{a}{3$\sigma$ uppper limits are estimated for non-detection.}
\end{deluxetable*}

\section{Discussion}\label{Discussion}
\subsection{Water abundance}\label{Water abundance}
We assume the initial gas surface density profile to follow a tapered power-law distribution \citep{1974_lynden-bell, 1998_hartmann, 2020_Harsono}:
\begin{equation}
\Sigma_\mathrm{g} (r)=\Sigma_\mathrm{c}\left(\frac{r}{r_{c}}\right)^{-\gamma}\exp\left[-\left(\frac{r}{r_{c}}\right)^{2-\gamma}\right]
\end{equation}
where $r_\mathrm{c}=75$ au \citep{2018_vanthoff,2021_leemker}, $\gamma=1.25$ \citep{2024_houge}, the disk mass is assumed to be
$M_\mathrm{disc} \approx 0.3 M_\odot$ \citep{2018_cieza,2021_leemker}, and
\begin{equation}
\Sigma_{c}=\frac{M_\mathrm{disc}(2-\gamma)}{2\pi r_{c}^{2}}.
\end{equation}

This method to estimate the initial gas surface density profile is based on \citet{2020_Harsono}.
Using the derived surface density profile, we calculate the $\mathrm{H}_{2}$ column density as
\begin{equation}
N_\mathrm{H_{2}}=\frac{\Sigma_\mathrm{g}}{\mu_\mathrm{H_2}m_\mathrm{H}}
\end{equation}
with $\mu_\mathrm{H_{2}}=2.8$ \citep{2008_kauffmann}.
At $r = 80$ au, corresponding to the water snowline, we obtain 
$N_\mathrm{H_{2}} = 3.9\times10^{24}\,\mathrm{cm^{-2}}$.

The H$_{2}$O total column densities were estimated from the H$_{2}$$^{18}$O column densities listed in Table~\ref{tab:column_density}, detected in Bands 5 and 7, by adopting the corresponding isotopic ratio 
($^{16}\mathrm{O}/^{18}\mathrm{O} = 560 \pm 25$) \citep{1994_wilson}.
Based on these values, we derived the water vapor abundances with respect to molecular hydrogen, expressed as $N_{\mathrm{H_{2}O}}/N_{\mathrm{H_2}}$, for excitation temperatures of $T_\mathrm{ex}=75,\ 180,\ \mathrm{and}\ 225\ \mathrm{K}$.
The results are summarized in Table~\ref{tab:abundance}.
The derived values of water vapor abundance are between $3\times10^{-7}$--$5\times10^{-6}$.
The variation in the abundance trends between Bands 5 and 7 with different assumed excitation temperatures $T_\mathrm{ex}$ is attributable to the differences in the upper state energies $E_\mathrm{u}$ of the H$_2$$^{18}$O lines in each band, as discussed in Section~\ref{seq:total column density}.

If water molecules are mostly inherited from dark clouds and pre-stellar cores, the water abundance should be $\sim 10^{-4}$ (e.g., \citealt{2015_Boogert}).
However, recent water line observations toward several Class 0 low-mass protostars suggest low water gas fractional abundances ($\sim 10^{-7}$--$10^{-6}$) in the inner warm envelopes ($r < 10^{2}\ \mathrm{au}$, within its water snowline, e.g., \citealt{2012_Persson, 2016_Persson, 2021_vanDishoeck, 2025_vanDishoeck}).
\citet{2021_Notsu} suggested that X-ray-induced destruction processes of water (ion-molecule reactions and the X-ray-induced photodissociation) can explain the observed lower water abundances in the inner envelopes of these objects.

In addition, \citet{2020_Harsono} conducted water line observations with ALMA and NOEMA and showed that water vapor is not abundant in the warm envelopes and disks around Class I protostars, and upper limit values of the water gas abundance averaged over the inner warm disk with $> 100\ \mathrm{K}$ are $\sim 10^{-7}$--$10^{-5}$.
These lower water gas abundances might also be caused by efficient water gas destruction through X-ray-induced chemistry, in addition to locking up water in icy dust grains (see also \citealt{2021_Notsu}).
Our derived values of water vapor abundances for the V883 Ori disk fall within this range, indicating that the water abundances of the V883 Ori disk inferred from our observations are consistent with previous studies of Class I and several Class 0 protostellar envelopes and disks.

\begin{deluxetable*}{lccc}
\tablewidth{0pt}
\tablecaption{Water abundances at different excitation temperatures $T_\mathrm{ex}$\label{tab:abundance}}
\tablehead{
\colhead{Band} &
\multicolumn{3}{c}{$\log(N_\mathrm{H_{2}O}/N_{H_2})$}
\\
&
\colhead{$(T_\mathrm{ex} = 75\ \mathrm{K})$} &
\colhead{$(T_\mathrm{ex} = 180\ \mathrm{K})$} &
\colhead{$(T_\mathrm{ex} = 225\ \mathrm{K})$}
}
\startdata
\textbf{Band 5} & $-5.93 \pm 0.08$ & $-6.06 \pm 0.08$ & $-6.02 \pm 0.08$ \\
\hline
\textbf{Band 7} & $-5.39 \pm 0.08$ & $-6.40 \pm 0.08$ & $-6.49 \pm 0.08$ \\
\enddata
\end{deluxetable*}

\subsection{HDO/H$_2$O ratio}
\begin{figure*}[htb!]
\centering
\includegraphics[width=0.6\textwidth]{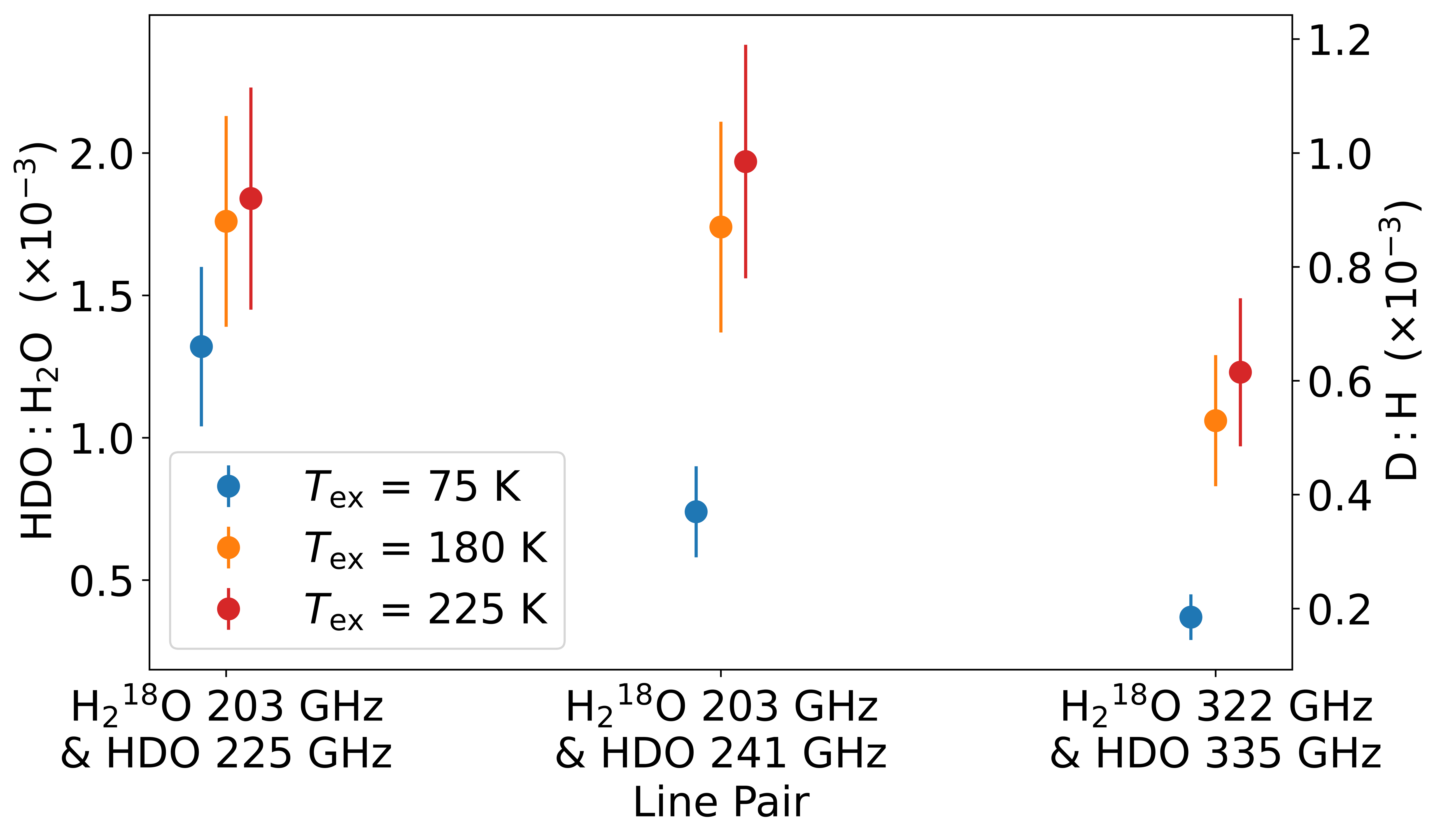}
\caption{\label{fig:hdo_h2o_ratio}
HDO/H$_2$O abundance ratios derived from pairs of water isotopologue lines observed in Band 5, 6, and 7 at excitation temperatures of 75, 180, and 225 K.   
Vertical error bars represent 1 standard deviation uncertainties for each data point.
}
\end{figure*}

We derived the HDO/H$_2$O abundance ratios using total column densities calculated from observed line pairs.
The H$_2$$^{18}$O column densities were converted to H$_2$$^{16}$O by applying the isotopic ratio $(^{16}\mathrm{O}/^{18}\mathrm{O} = 560 \pm 25)$ \citep{1994_wilson}, enabling a direct comparison between HDO and the main isotopologue of water.
The D/H ratios were calculated using the relation $\mathrm{HDO/H_2O} = 2\times\mathrm{D/H}$. 
The analysis was performed assuming excitation temperatures of 75 K, 180 K, and 225 K, following the same assumptions as \citet{2023_tobin}, and the results were summarized for each line pair in Table~\ref{tab:hdo_h2o_ratio} and Figure~\ref{fig:hdo_h2o_ratio}.
Although in our analyses the HDO/H$_2$O abundance ratios estimated from line pairs observed in Band 7 tend to be lower than those derived from line pairs observed in Bands 5 and 6 (see Table~\ref{tab:hdo_h2o_ratio} and Figure~\ref{fig:hdo_h2o_ratio}),  
these results do not contradict the range of ratios reported in previous studies \citep{2023_tobin}. 

The abundance of deuterated species becomes enhanced in the cold ISM due to deuterium becoming locked into $\mathrm{H}_{2}\mathrm{D}^{+}$, and the dissociative recombination of deuterated molecules increases the free deuterium available for HDO and D$_{2}$O formation within ice mantles of dust grains \citep{2016_furuya, 2017_furuya, 2021_vanDishoeck, 2023_nomura}.
In contrast, within the warm gas in the inner envelopes around Class 0 protostars, where ices have sublimated, the ratios have been measured more precisely with ALMA and NOEMA and found to lie roughly between \(6 \times 10^{-4}\) and \(2 \times 10^{-3}\) (e.g., \citealt{2014_persson, 2019_jensen, 2021_jensen, 2020_Harsono, 2021_Notsu, 2023_tobin, 2025_vanDishoeck}).
Recently, the HDO/H$_2$O ice ratios in low-mass protostars are reported as $\sim5\times10^{-3}$ from JWST spectra \citep{2025_slavicinska, 2025_vanDishoeck}.
The HDO/H$_2$O in comets range between $10^{-4}$ and $10^{-3}$ (e.g., \citealt{2015_altwegg, 2019_altwegg, 2023_nomura, 2025_cordiner}).
The HDO/H$_2$O abundance ratios of the V883 Ori disk reported in our analyses and other studies \citep{2023_tobin, 2025_leemkera} are comparable to the values reported in Class 0 protostars for water vapor and ice, and also similar to the values for many Oort Cloud comets and 67P/Churyumov-Gerasimenko from the Jupiter Family System (e.g., \citealt{2015_altwegg, 2019_altwegg, 2023_tobin}), suggesting that water in the V883 Ori would be directly inherited from the infalling protostellar envelopes \citep{2016_furuya, 2017_furuya, 2021_vanDishoeck}.

\begin{deluxetable*}{lccc}
\tablewidth{0pt}
\tablecaption{Derived HDO/H$_2$O abundance ratios at different excitation temperatures\label{tab:hdo_h2o_ratio}}
\tablehead{
\colhead{Line Pair} &
\multicolumn{3}{c}{HDO/H$_{2}$O}
\\
&
\colhead{$(T_\mathrm{ex} = 75\ \mathrm{K})$} &
\colhead{$(T_\mathrm{ex} = 180\ \mathrm{K})$} &
\colhead{$(T_\mathrm{ex} = 225\ \mathrm{K})$}
}
\startdata
H$_2$$^{18}$O (203 GHz), HDO (225 GHz) & $(1.32\pm0.28)\times10^{-3}$ & $(1.76\pm 0.37)\times10^{-3}$ & $(1.84\pm0.39)\times10^{-3}$ \\
H$_2$$^{18}$O (203 GHz), HDO (241 GHz) & $(0.74\pm0.16)\times10^{-3}$ & $(1.74\pm0.37)\times10^{-3}$ & $(1.97\pm0.41)\times10^{-3}$ \\
H$_2$$^{18}$O (322 GHz), HDO (335 GHz) & $(0.37\pm0.08)\times10^{-3}$ & $(1.06\pm0.23)\times10^{-3}$ & $(1.23\pm0.26)\times10^{-3}$ \\
\enddata
\end{deluxetable*}

\subsection{Band 7 HDO Line Weakness Relative to Rotational Diagram Predictions}\label{seq: Band 7 HDO Line Weakness Relative to Rotational Diagram Predictions}
Section~\ref{Rotation Temperature and Column Density} demonstrated that the Band 7 HDO transition at 335 GHz falls well below the trend defined by the Band 6 lines, with an observed intensity roughly a quarter of the expected strength.
Here, we examine possible causes for this discrepancy.

\subsubsection{Possibility of a more compact emitting region}\label{seq: Possibility of a more compact emitting region}
A comparison of the upper state energies ($E_\mathrm{u}$) of the detected HDO transitions reveals that the Band 6 lines at 225 GHz and 241 GHz (with $E_\mathrm{u} = 167$ K and 95 K, respectively) are excited under significantly cooler conditions than the Band 7 transition at 335 GHz $(E_\mathrm{u} = 335\ \mathrm{K})$.
This suggests that the 335 GHz line could originate in warmer regions, likely closer to the central protostar.

Under the assumption that line intensity scales approximately with the emitting area, the relatively weak intensity of the 335 GHz transition implies a more compact emission region.
Assuming the same inner-disk obscuration assumption used in the Band 6 analysis—--i.e., that the inner 40 au is heavily obscured by dust and contributes little to the observed emission—--and adopting $\sim80\ \mathrm{au}$ as the emission radius for the Band 6 HDO line \citep{2023_tobin}, we estimate an effective emission radius of about 53 au for the Band 7 HDO line.

This interpretation is consistent with previous theoretical studies \citep{2016_notsu, 2017_notsu, 2018_notsu}, which argue that water lines with relatively low Einstein A coefficients and high excitation energies serve as effective tracers of hot water vapor inside the snowline, where dust temperatures are sufficiently high to enable thermal desorption.
Transitions such as para-H$_2$$^{18}$O $5_{1,5}$–$4_{2,2}$ at 322 GHz and HDO $3_{3,1}$–$4_{2,2}$ at 335 GHz (along with undetected lines like ortho-H$_2$$^{16}$O $10_{2,9}$–$9_{3,6}$ at 321 GHz) have been identified as valuable tracers of warm water in the inner disk.

Therefore, our conclusion that the 335 GHz HDO emission arises from a more compact, warmer region than that traced by the Band 6 lines is not only observationally supported, but also theoretically plausible.

\subsubsection{Possibility of a larger central depression due to dust optical depth}
Dust opacity can significantly attenuate molecular line emission from the innermost regions, as previously demonstrated for this source \citep{2018_vanthoff, 2019_lee, 2024_yamato}.  
Furthermore, \citet{2020_desimone} found that dust absorption at millimeter wavelengths can substantially reduce the intensity of optically thin lines, leading to underestimates of molecular abundances and an apparent deficiency in the detected COMs.  
In (sub)millimeter observations, bright dust continuum emission can often obscure molecular lines, especially for species residing in the disk midplane.  
In the case of V883 Ori, the dust optical depth at the disk midplane ($r \sim 42$ au) could exceed $\tau_\mathrm{dust} \gtrsim 2$ \citep{2016_cieza}, potentially causing significant underestimates of molecular line intensities from this region.

\citet{2025_jeong} estimated the averaged dust optical depth, $\tau_{\nu,\mathrm{dust}}$, over the COM-emitting region, $\sim0\farcs1$--$0\farcs3$ \citep{2024_lee}, using the following equation:
\begin{equation}\label{eq:2}
\tau_{\nu,\ \mathrm{dust}}
=
(N_\mathrm{H_{2},cavity} \cdot \kappa_\mathrm{ref} \cdot m_\mathrm{H_{2}} \cdot 0.01)
\cdot
\left(
\frac{\nu}{\nu_\mathrm{ref}}
\right)^{\beta}
\end{equation}

In this study, we adopt the same approach, using observational parameters derived from Band 6 data.  
We assume a dust mass opacity $\kappa_\mathrm{ref} = 2.2~\mathrm{cm^2~g^{-1}}$ at $\nu_\mathrm{ref} = 230$ GHz and a dust opacity spectral index $\beta = 1$, following \citet{2018_cieza}.  
We also adopt the molecular hydrogen column density $N_\mathrm{H_{2},\ cavity} = 1.24 \times 10^{25}\ \mathrm{cm^{-2}}$ as derived by \citet{2025_jeong}.

By substituting these values into Equation~(\ref{eq:2}), we estimate $\tau_{\nu,\ \mathrm{dust}}$ and the corresponding line attenuation due to dust extinction ($e^{-\tau_{\nu,\ \mathrm{dust}}}$) for the three isotopologue water lines reported in \citet{2023_tobin} and the two lines detected in this study.  
The results are summarized in Table~\ref{tab:tau_dust}.

\begin{deluxetable}{lcc}
\tablewidth{0pt}
\tablecaption{$\tau_{\nu,\ \mathrm{dust}}$, $e^{-\tau_{\nu,\ \mathrm{dust}}}$\label{tab:tau_dust}}
\tablehead{
\colhead{Isotope (Frequency)} &
\colhead{$\tau_{\nu,\ \mathrm{dust}}$} &
\colhead{$e^{-\tau_{\nu,\ \mathrm{dust}}}$}
}
\startdata
\multicolumn{3}{l}{\textbf{Band 5}} \\
\hline
H$_2$$^{18}$O (203 GHz) & 0.80 & 0.45 \\
\hline
\multicolumn{3}{l}{\textbf{Band 6}} \\
\hline
HDO (225 GHz) & 0.89 & 0.41 \\
HDO (241 GHz) & 0.96 & 0.39 \\
\hline
\multicolumn{3}{l}{\textbf{Band 7}} \\
\hline
H$_2$$^{18}$O (322 GHz) & 1.28 & 0.28 \\
HDO (335 GHz) & 1.33 & 0.27 \\
\enddata
\end{deluxetable}

Based on our calculations, the water lines detected in Band 7 are subject to dust attenuation somewhat stronger than those detected in Bands 5 and 6, by a factor of roughly 1.4--1.6.
This qualitatively agrees with our observational result that the Band 7 HDO line intensities are weaker than predicted.
However, quantitatively, the observed Band 7 HDO lines are only about one-quarter of the predicted values.
To quantitatively explain the weaker Band 7 water line compared to the Band 6 lines by the stronger dust attenuation, one potential approach is to attribute it to the difference in sizes of the dust-induced cavity regions in the disk center reported by \citet{2023_tobin}.
If the reduced intensity of the HDO line observed in Band 7 is attributed to the frequency-dependent impact of the dust-induced cavity, this implies that the apparent cavity radius is $\sim40\ \mathrm{au}$ at Band 6 but increases to $\sim60\ \mathrm{au}$ at Band 7.
However, the spatial resolution of the observations presented in this study is insufficient to resolve such cavities clearly.

\subsubsection{Future Prospects}
As discussed above, the significant reduction in the observed Band 7 HDO line intensity compared with the Band 6 HDO line can be interpreted as arising from (i) a more compact emitting region due to the higher upper state energy transition, (ii) an expansion of the apparent cavity caused by dust continuum absorption affecting the line emission, or (iii) a combination of both effects.
Considering that the excitation temperature of the Band 7 H$_2$$^{18}$O line is also estimated to be relatively low, its weaker intensity compared with the Band 5 H$_2$$^{18}$O line may similarly be attributed to the same factors affecting the Band 7 HDO line.

However, the spatial resolution of the observations presented in this study is insufficient to directly resolve the extent of the emitting region or to clearly identify the apparent cavity structure due to dust optical depth, as illustrated in Figure~\ref{fig:moment0_hdo} and~\ref{fig:moment0_h218o}.
Therefore, to definitively determine the cause of the significantly lower Band 7 HDO intensity compared to the value predicted from Band 6 observations \citep[e.g.,][]{2023_tobin}, observations with higher spatial resolution will be indispensable.

\section{Summary}\label{Summary}
We observed the protoplanetary disk around the FU Ori-type star V883 Ori with ALMA Band 7 and detected two targeted transitions of water: para-H$_2$$^{18}$O $5_{1,5}$–$4_{2,2}$ at 322 GHz and HDO $3_{3,1}$–$4_{2,2}$ at 335 GHz.
By carefully removing blends with nearby molecular lines, we estimated the intrinsic line fluxes of the detected water transitions.
With the correction for Keplerian rotation applied, HDO and H$_2$$^{18}$O were detected with significances of 23.6$\sigma$ and 9.3$\sigma$, respectively.
Combining these Band 7 detections with previously reported water lines in Band 5 and 6, we constructed rotational diagrams to derive the column densities and rotational temperatures of each water isotopologue.
From the derived column densities, we estimated water vapor abundances relative to H$_2$ at the water snowline ($\sim80\ \mathrm{au}$), obtaining values of $3\times10^{-7}$--$5\times10^{-6}$.
Note that our derived water abundance value is the averaged value in the vertical direction.
These abundances are consistent with the low water gas fractions reported for Class 0 and I protostellar envelopes and disks, which may be explained by X-ray induced destruction and/or water locked in icy grains.
We also derived HDO/H$_2$O abundance ratios in the range (0.4--2.0)$\times10^{-3}$, comparable to values measured in gas and ice in Class 0 protostellar envelopes and comets, suggesting that water in the V883 Ori disk is largely inherited from the parental protostellar envelope.

We found that the Band 7 HDO lines are significantly weaker than expected based on the rotational diagram derived from the Band 6 transitions.
This discrepancy suggests that the Band 7 emission could arise from a more compact and narrow region than the Band 6 lines.
Such a compact emitting region may result from the higher upper state energies of the Band 7 lines, which trace hotter gas located closer to the central protostar, and/or from higher optical depths, which would make the central dust cavity appear larger than in the Band 6 observations due to stronger dust absorption.
Taken together, our analysis indicates that the V883 Ori disk shows water abundances and isotopic ratios consistent with inheritance from protostellar envelopes, while the weakness of the Band 7 HDO lines points to more compact or optically thick emission.
To distinguish between these possibilities and to constrain the precise spatial distribution of the Band 7 water emission, higher angular resolution follow-up observations are needed.

\begin{acknowledgments}
The authors thank the anonymous referee for comments.
We are grateful to Professor J. J. Tobin for valuable suggestions on the analysis code and to Dr. Nanase Harada for useful comments.
We are also grateful to the summer student program at the National Astronomical Observatory of Japan / Astronomical Science Program of SOKENDAI, where this research initially took place.
This paper makes use of the following ALMA data: ADS\/JAO.ALMA\#2021.1.00115.S and \#2021.1.00186.S.
ALMA is a partnership of ESO (representing its member states), NSF (USA), and NINS (Japan), together with NRC (Canada), MOST, ASIAA (Taiwan), and KASI (Republic of Korea), in cooperation with the Republic of Chile.
The Joint ALMA Observatory is operated by ESO, AUI/NRAO, and NAOJ.
ALMA Data analysis was carried out on the Multi-wavelength Data Analysis System operated by the Astronomy Data Center (ADC), National Astronomical Observatory of Japan.
H.N was supported by the ALMA Japan Research Grant of NAOJ ALMA Project, NAOJ-ALMA-374.
S.N. is grateful for support from Grants-in-Aid for JSPS (Japan Society for the Promotion of Science) Fellows grant No. JP23KJ0329, MEXT/JSPS Grants-in-Aid for Scientific Research (KAKENHI) grant Nos. JP23K13155, JP24K00674, and JP23H05441, and Start-up Research Grant as one of The University of Tokyo Excellent Young Researcher 2024.
T.H. is financially supported by the MEXT/JSPS KAKENHI grant Nos. 17K05398, 18H05222, and 20H05845.
T.C.Y. thanks for support from Grants-in-Aid for JSPS Fellows grant No. JP23KJ1008.

\facility{
ALMA
}

\software{
CASA \citep{2022_casateam},
Astropy \citep{2013_astropycollaboration, 2018_astropycollaboration, 2022_astropycollaboration},
GoFish \citep{2019_teague},
Numpy \citep{2020_harris},
Scipy \citep{2020_virtanen},
Matplotlib \citep{2007_hunter},
Emcee \citep{2013_foreman-mackey},
corner.py \citep{2016_foreman-mackey}
}

\end{acknowledgments}

\appendix

\section{Disk-integrated Spectra}\label{app:appendixA}

\begin{figure*}[htb!]
\centering
\includegraphics[width=0.7\textwidth]{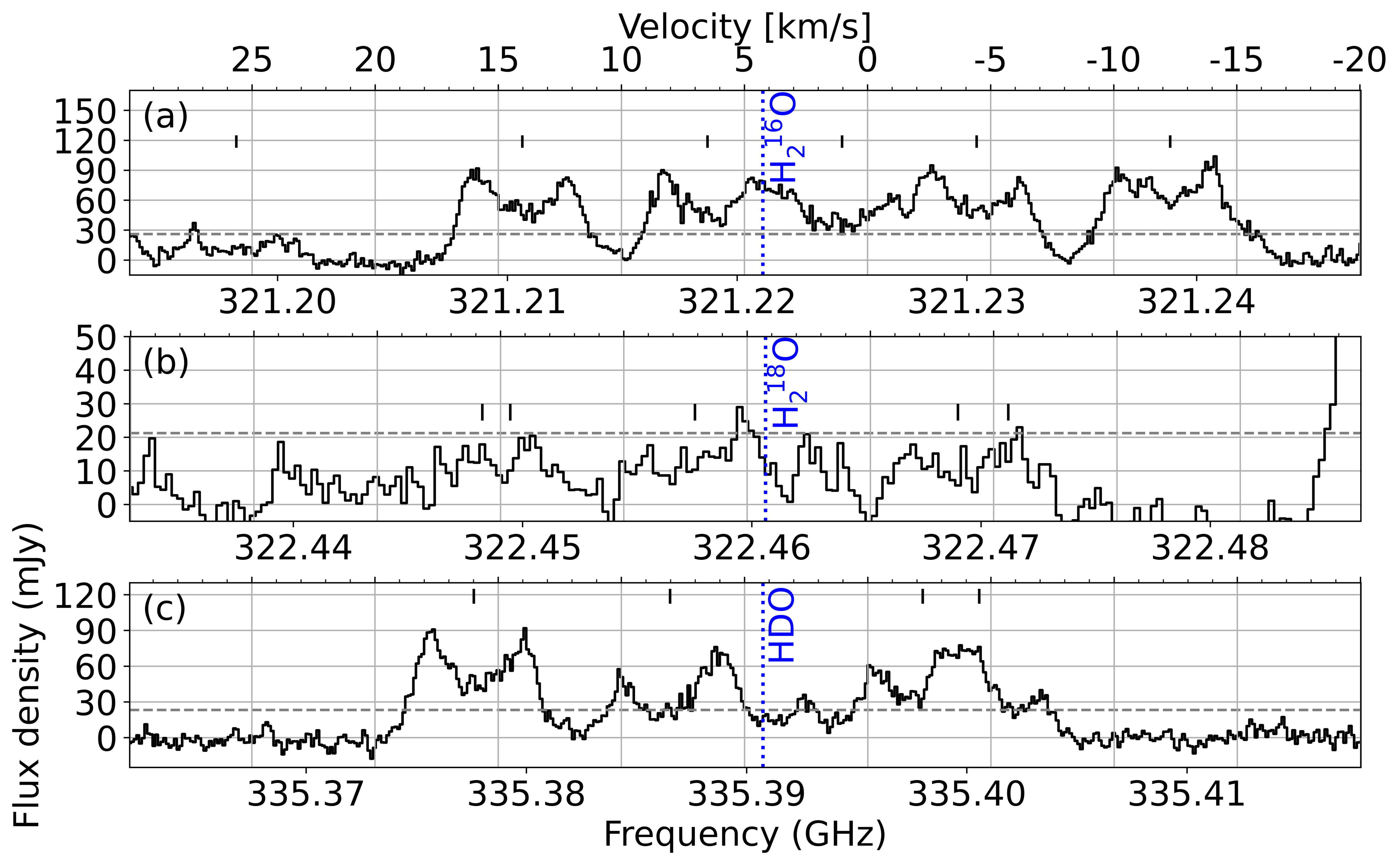}
\caption{\label{fig:spectra_unstack} Disk-integrated spectra for the three spectral windows targeting the H$_2$$^{16}$O 321 GHz, H$_2$$^{18}$O 322 GHz, and HDO 335 GHz lines (from top to bottom).
The spectra exhibit double-peaked profiles due to Keplerian rotation.
Short black lines above the spectra indicate the frequencies of detected contaminated lines.}
\end{figure*}

\begin{figure*}[htb!]
\centering
\includegraphics[width=0.7\textwidth]{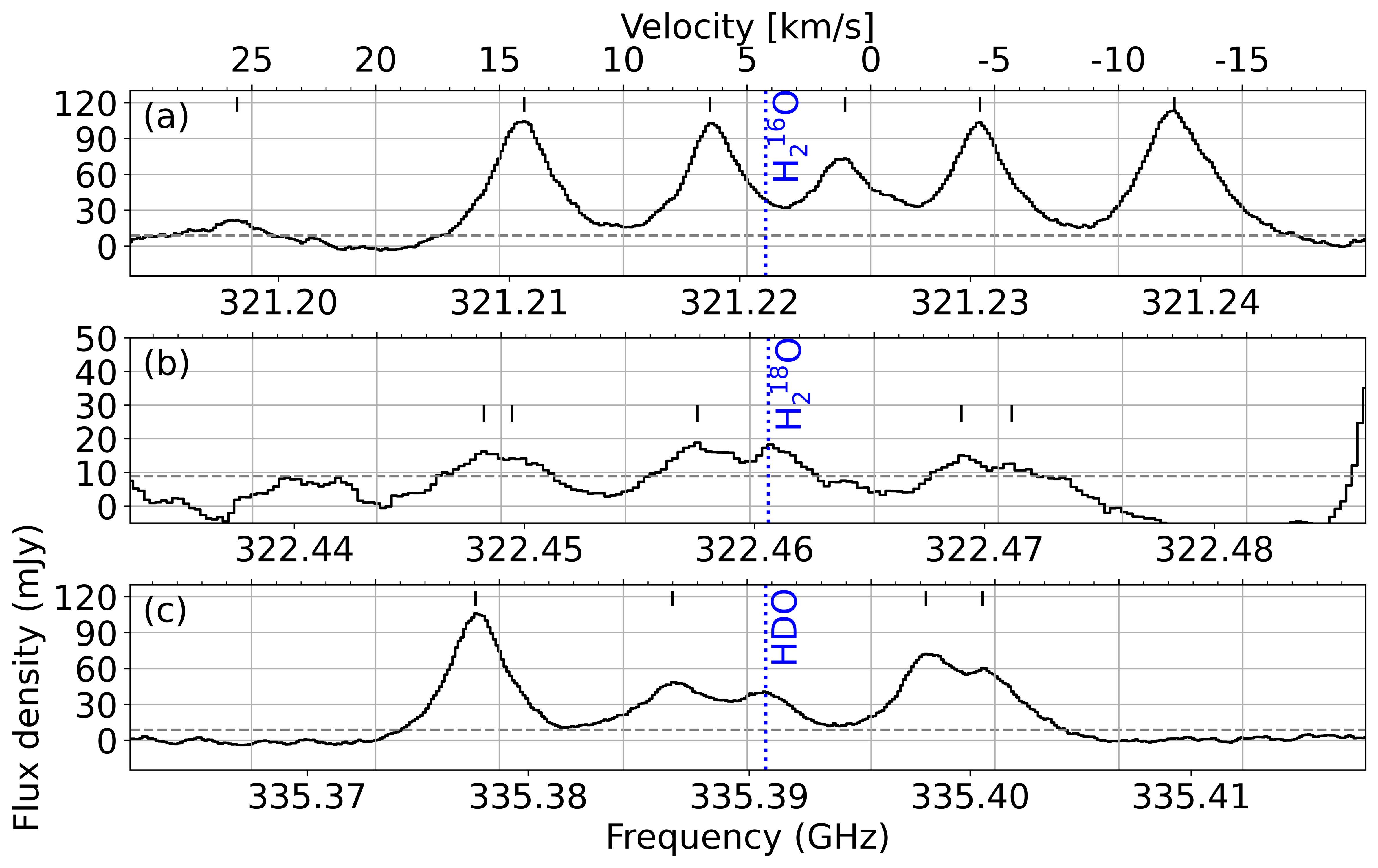}
\caption{\label{fig:spectra_stack} Velocity-aligned spectra of the same lines as in Figure~\ref{fig:spectra_unstack}, obtained using \texttt{GoFish}.
The alignment removes rotational broadening and yields single-peaked profiles.
Short black lines above the spectra mark the rest frequencies of detected contaminated lines.}
\end{figure*}

Figure~\ref{fig:spectra_unstack} displays the disk-integrated spectrum for each spectral window.
The observed line profiles exhibit characteristic double-peaked shapes, indicative of Keplerian rotation.
To recover the intrinsic line shapes, we applied spectral alignment using \texttt{GoFish}, which effectively removes the velocity gradients introduced by rotation.
The aligned spectra (Figure~\ref{fig:spectra_stack}) exhibit single-peaked profiles centered around the systemic velocity.
Short black lines above the spectra indicate the frequencies of detected contaminated lines.
The corresponding rest frequencies are summarized in Table~\ref{tab:kep_mask}, and these values were used to create the Keplerian masks shown in Figures~\ref{fig:channel_hdo} and~\ref{fig:channel_h218o}.
The expected positions of the targeted water lines at the systemic velocity are marked with blue dashed lines.

\section{Comparison of Model Spectra for Fitting}\label{app:appendixB}
The spectral fitting was performed using template spectra constructed from molecular lines detected in the observational data.
Specifically, we employed two types of template spectra: one based on the CH$_3$OH 242 GHz lines detected in Band 6, as used in \citet{2023_tobin} (the Band 6 template), and the other based on the CH$_3$CHO 335 GHz line detected in our Band 7 data (the Band 7 template).

\begin{figure*}[htb!]
\centering
\includegraphics[width=0.7\textwidth]{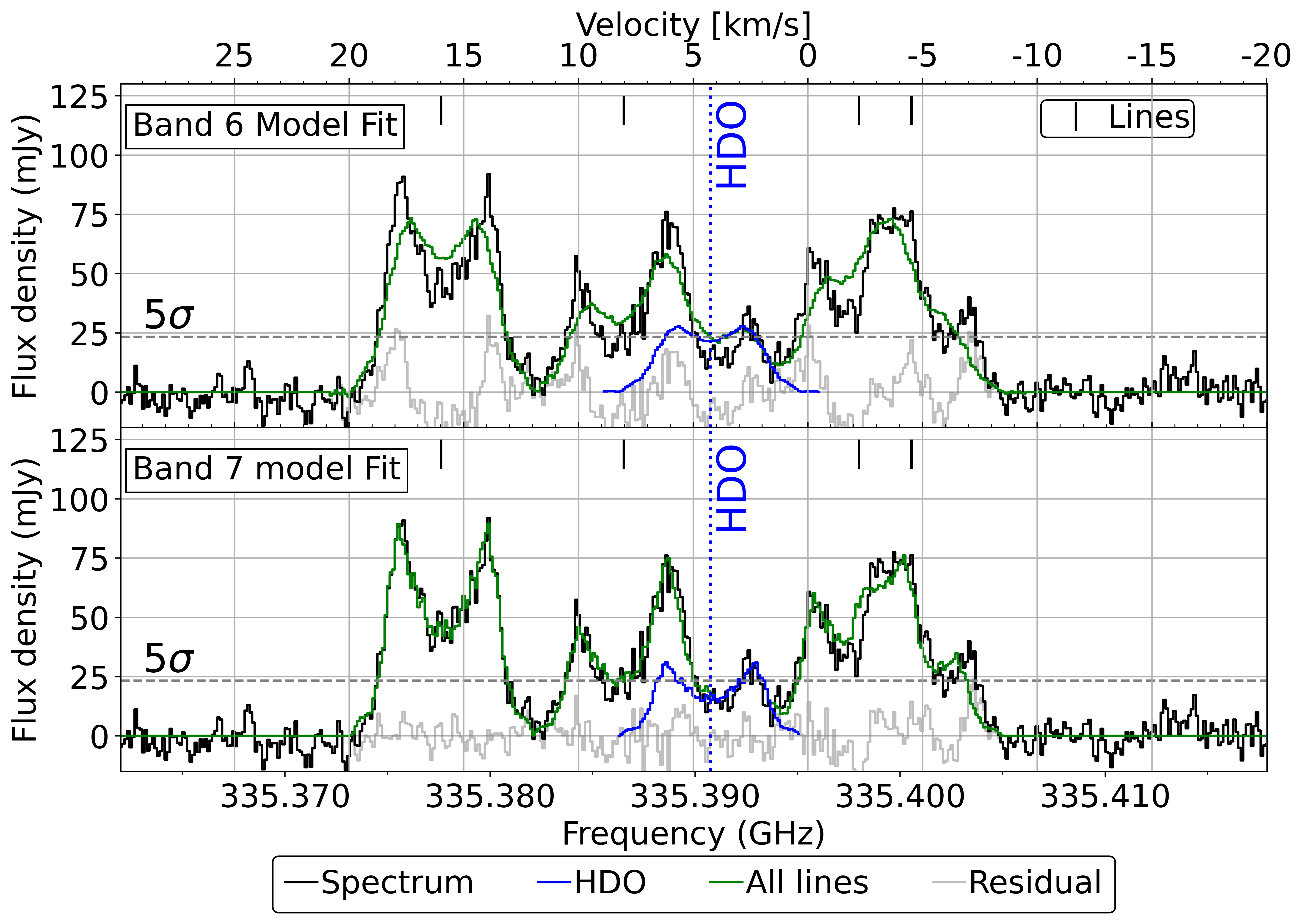}
\caption{\label{fig:model}
Spectra around the HDO 335 GHz emission line observed in Band 7, with fitting results based on model spectra derived from Band 6 (upper panels) and Band 7 (lower panels) data.
The black solid lines represent the observed spectra.
The blue dotted line indicates the frequency of the HDO line corresponding to the systemic velocity of V883 Ori (4.25 km s$^{-1}$), while the blue solid lines show the fitted models for the HDO line.
The green solid lines represent the combined fitting models including the HDO line and contaminated lines.
The gray solid lines represent the residuals obtained by subtracting the model spectrum from the observed data.
Vertical black ticks indicate the central frequencies of the contaminated lines, and the gray dashed lines correspond to the 5$\sigma$ noise level.
The fitting using the template derived from Band 7 spectrum achieves a more accurate reproduction of the observed features compared to that using the Band 6 template.
}
\end{figure*}

Using both templates, we fitted the spectra to derive the integrated intensities (see Figure~\ref{fig:model}) and their uncertainties for the water isotopologue lines.
The uncertainties were estimated from the residuals in the frequency range corresponding to each target line.

For all water isotopologue lines in Bands 5, 6, and 7, the uncertainties were systematically smaller when using the Band 7-based template (see Table~\ref{tab:template}).
This difference may arise from the fact that the lines used to construct the Band 6 template could be optically thick, resulting in a different line profile.
Therefore, we adopted the fitting results from the Band 7 template for the subsequent calculations of column densities and excitation temperatures.

We also compared the line profiles of the HDO lines detected in Band 5 and 6 by \citet{2023_tobin} with those of the Band 7 HDO line detected in this work, in addition to the model profiles.
As a result, we did not find any significant difference in the peak positions of the lines.
The line profiles correspond to the emitting regions of each transition \citep{2016_notsu, 2017_notsu, 2018_notsu}.
Assuming the emitting region of 53 au estimated in Section~\ref{seq: Possibility of a more compact emitting region}, the difference between this size and the emitting region of 80 au inferred from the Band 7 observations corresponds to only a factor of 1.5 in radius, or a factor of 1.2 in the corresponding Keplerian velocity.
Therefore, given the spectral resolution of each observation, the lack of a clear difference in the peak positions for this low-mass object is hardly avoidable.

\begin{deluxetable}{lcc}
\tablewidth{0pt}
\tablecaption{Integrated intensities derived from fitting using Band 6 and Band 7 template spectra $I_\mathrm{int}$ \label{tab:template}}
\tablehead{
\colhead{Line} & 
\colhead{$I_\mathrm{int,\,Band\,6}$} &
\colhead{$I_\mathrm{int,\,Band\,7}$} \\
\colhead{} &
\colhead{(mJy km s$^{-1}$)} &
\colhead{(mJy km s$^{-1}$)}
}
\startdata
\hline
\multicolumn{3}{l}{\textbf{Band 5}}\\
\hline
H$_2$$^{18}$O (203 GHz) & $168.51 \pm 23.78$ & $144.52 \pm 20.75$ \\
\hline
\multicolumn{3}{l}{\textbf{Band 6}}\\
\hline
HDO (225 GHz) & $588.70 \pm 24.37$ & $585.92 \pm 22.14$ \\
HDO (241 GHz) & $528.00 \pm 35.32$ & $555.59 \pm 32.82$ \\
\hline
\multicolumn{3}{l}{\textbf{Band 7}}\\
\hline
H$_2$$^{16}$O (321 GHz) & - & - \\
H$_2$$^{18}$O (322 GHz) & $\phantom{0}55.96 \pm 8.01$ & $\phantom{0}52.06 \pm 7.48$ \\
HDO (335 GHz) & $133.00 \pm 9.37$ & $123.73 \pm 6.15$ \\
\enddata
\tablecomments{
$I_\mathrm{int,\,Band\,6}$ and $I_\mathrm{int,\,Band\,7}$ were derived by fitting template spectra constructed from detected contaminated lines in Band 6 and Band 7, respectively.
For H$_2$$^{16}$O (321 GHz), the line was not detected.
The $3\sigma$ upper limit is listed in Table~\ref{tab:line_flux}.
}
\end{deluxetable}

\section{Rotational Diagram of HDO in ALMA Band 6}\label{app:appendixC}
\begin{figure*}[htb!]
\centering
\begin{minipage}[b]{0.48\textwidth}
  \centering
  \includegraphics[width=0.8\textwidth]{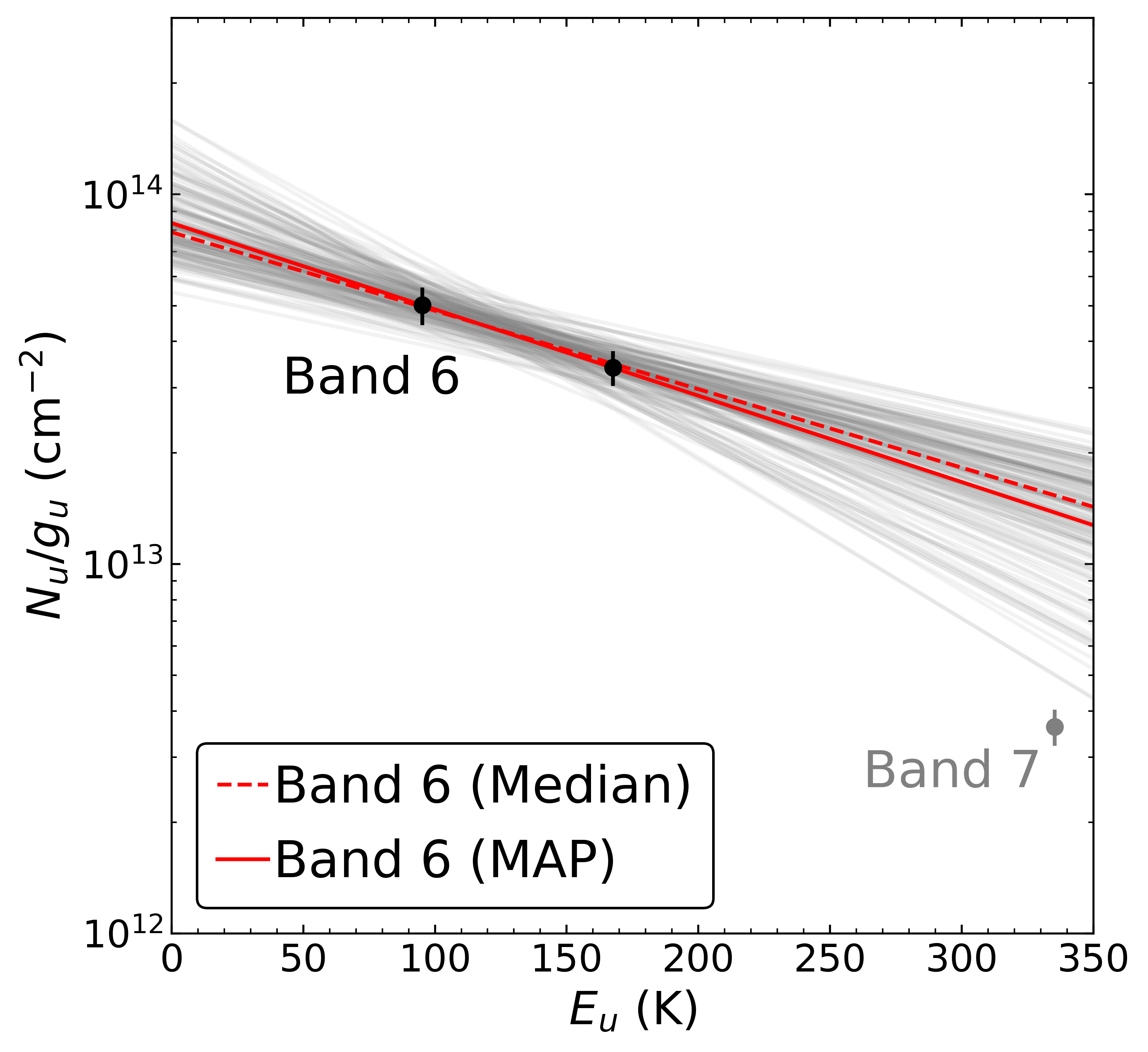}
\end{minipage}
\hfill
\begin{minipage}[b]{0.48\textwidth}
  \centering
  \includegraphics[width=0.8\textwidth]{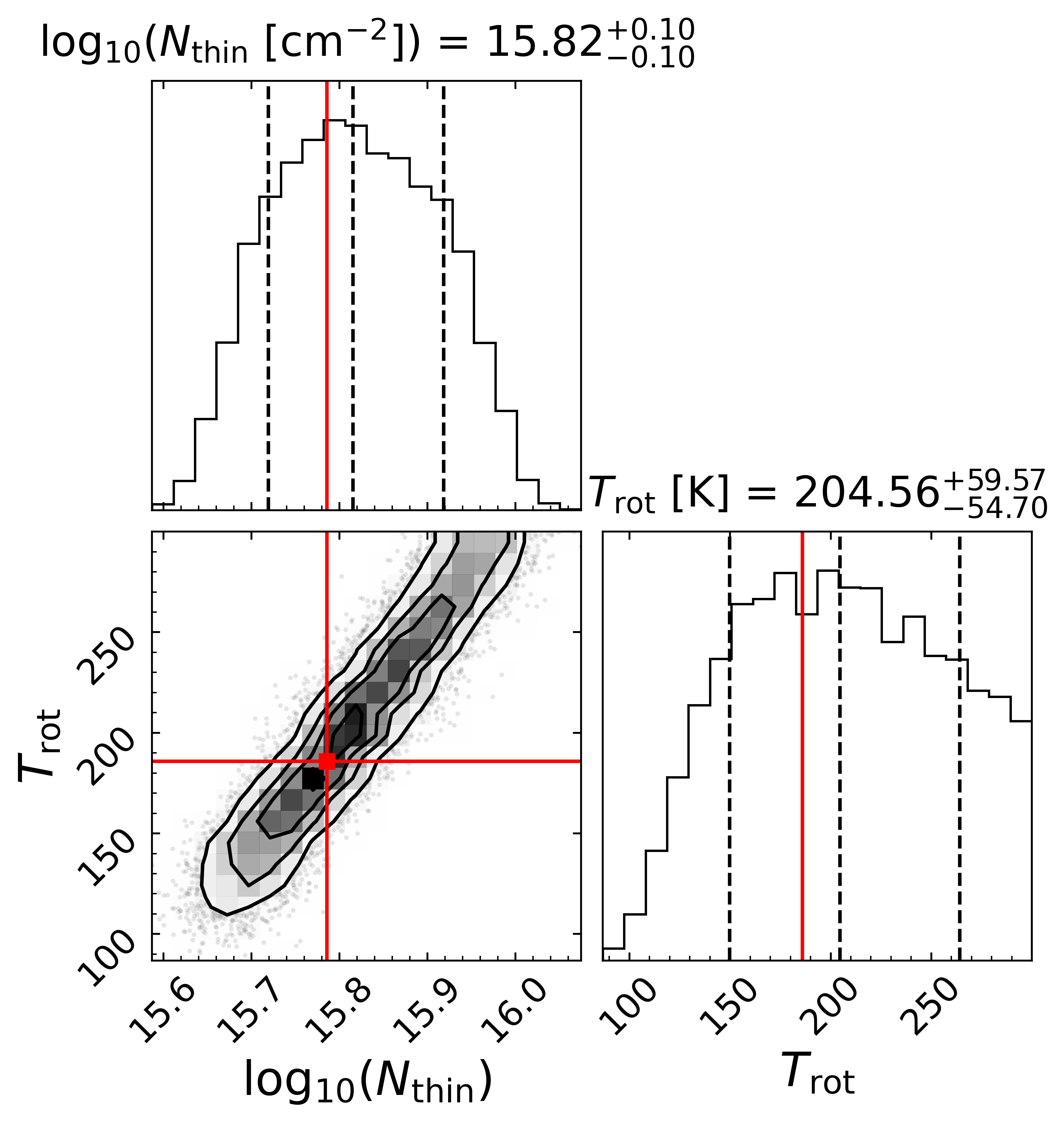}
\end{minipage}
\caption{\label{fig:rotation_hdo_band6}%
The same as Figure~\ref{fig:rotation_hdo}, but for two HDO lines observed with ALMA Band 6.}
\end{figure*}

Figure~\ref{fig:rotation_hdo_band6} shows the rotational diagram constructed from the two Band 6 HDO lines reported by \citet{2023_tobin}.
The red solid line in the figure represents the MAP estimate from our MCMC analysis, which corresponds to the red solid line shown in Figure~\ref{fig:rotation_hdo}.
The results of the MAP estimation indicate that HDO has a rotational temperature of 186.4 K and a logarithmic column density of $\log(N [\mathrm{cm}^{-2}]) = 15.79$.
As noted in Section~\ref{Spectral Fitting and Line Identification}, our treatment of the HDO line intensities differs from that of \citet{2023_tobin} in that we use the integrated fluxes of the fitted spectral models rather than direct flux measurements.
Nevertheless, the derived column densities and rotational temperatures are consistent within the 1$\sigma$ uncertainties.

\bibliography{00_reference}{}
\bibliographystyle{aasjournalv7}

\end{document}